\documentclass[12pt]{article}
\usepackage[margin=1.9cm]{geometry}
\usepackage{graphicx}
\usepackage{cite}

\newcommand{\mysection}{\setcounter{equation}{0}\section}

\def\beq{\begin{equation}}
\def\eeq{\end{equation}}
\def\beqa{\begin{eqnarray}}
\def\eeqa{\end{eqnarray}}
 
\begin{document}

\begin{center}
{\Large \bf Four-loop massive cusp anomalous dimension in QCD: \\ a calculation from asymptotics}
\end{center}

\vspace{2mm}

\begin{center}
{\large Nikolaos Kidonakis}\\

\vspace{2mm}

{\it Department of Physics, Kennesaw State University, \\
Kennesaw, GA 30144, USA}

\end{center}

\begin{abstract}
I present a general method for determining the massive cusp anomalous dimension in QCD to a very high degree of accuracy using its asymptotic behavior at small and large quark velocities. I show that the method works exceedingly well at two and three loops where exact results are already known. I then present a calculation of the massive cusp anomalous dimension using its asymptotics at four loops, and I provide a detailed study of the results for different values of the number of flavors and for separate color structures. The method can be further extended and applied to higher numbers of loops.
\end{abstract}

\mysection{Introduction}

The cusp anomalous dimension \cite{AMP,BNS,KnSc,IKR,KRsj,KR1,KR2,GPK,KR3,KMM,AGhq,NK2loop,NK2lp,NKtW,CHMS,GHKM1,GHKM2,NK3loopcusp,GHS,AGG1,BGHS,NK3loop,BDHY,GLP,AGG2} controls the infrared behavior of perturbative QCD scattering amplitudes. It is the simplest soft anomalous dimension in QCD and an essential ingredient of all calculations of soft anomalous dimensions for processes with more complicated color structures, see e.g. Refs. \cite{NKGS1,NKGS2,KOS,NKtt2l,FK2020}, and Ref. \cite{NKuni} for a review.

Wilson or eikonal lines describe the radiation of soft gluons by partons (i.e. quarks or gluons). The partons are represented by ordered exponentials in which the path is a straight line in the direction of the parton four-velocity $v$ as
\beq                                                   
W({\lambda}_2,{\lambda}_1;x)=
P\exp\left(-ig\int_{{\lambda}_1}^{{\lambda}_2}d{\lambda}\; 
v{\cdot} A ({\lambda}v+x)\right)\, ,
\label{Wl}
\eeq
where $P$ is an operator that orders group products in the same sense as ordering in the integration variable $\lambda$, and $A$ is the gauge field in the appropriate representation of the gauge group. The pattern of soft radiation is determined by the charge currents a long time before the scattering event and after it, which underlies the concept of factorization in QCD hard-scattering cross sections.
 
The cusp angle $\theta$ between two eikonal lines with four-velocities $v_1$ and $v_2$ is defined by the relation $\theta=\cosh^{-1}(v_1\cdot v_2/\sqrt{v_1^2 v_2^2})$. In simple processes such as $e^+ e^- \to t \, {\bar t}$, we have two eikonal lines meeting at a color singlet vertex. This vertex is associated with ultraviolet divergences which are dealt with renormalization. The anomalous dimension in the corresponding renormalization group equation is the cusp anomalous dimension, $\Gamma_{\rm cusp}$, and it is the same for all color singlets.

While the case of $\Gamma_{\rm cusp}$ with massless eikonal lines essentially involves only color coefficients and constants \cite{KT82,MVV04,HKM,MPS} and is known fully through four loops, the massive case has a complicated structure in terms of (harmonic) polylogarithms involving the masses of the eikonal lines \cite{NK2loop,NK2lp,GHKM1,GHKM2,NK3loopcusp} and is only known fully through three loops, with some terms as well as limits for small and large cusp angles known at four loops (see \cite{AGG2} for a recent review). 

We consider eikonal lines representing massive quarks that have the same mass $m$ and momenta $p_i^{\mu}=({\sqrt s}/2) v_i^{\mu}$, with $i=1,2$ and $s=(p_1+p_2)^2$, i.e. the case of production of a heavy quark-antiquark pair. Then, we have $v_1\cdot v_2=1+\beta^2$ and $v_1^2=v_2^2=1-\beta^2$, where $\beta=\sqrt{1-4m^2/s}$ is the quark speed. Then, the cusp angle is $\theta=\ln[(1+\beta)/(1-\beta)]$, and in reverse we have $\beta=\tanh(\theta/2)$. Clearly, the range of $\beta$ is from 0 (at absolute threshold with $s=4m^2$) to 1 (the massless case with $m=0$), and the corresponding range for $\theta$ is from zero to infinity.

The perturbative series for the cusp anomalous dimension in QCD is written as 
\beq
\Gamma_{\rm cusp}=\sum_{n=1}^{\infty} \left(\frac{\alpha_s}{\pi}\right)^n \Gamma^{(n)}
\eeq 
where $\alpha_s$ is the strong coupling. Beyond one loop, the expressions involve the number of light-quark flavors, $n_f$. We will show how to determine $\Gamma^{(n)}$ to a superb precision from its asymptotic behavior at large and small $\beta$. The two-loop and three-loop cases provide a stringent test of the method for all physical choices (and beyond) for $n_f$, and the method allows precise predictions at four loops. It is important to note that while we will derive results using the speed $\beta$, the results are not limited to the case of eikonal lines representing two quarks with the same mass. Once the method is used and the results are then reexpressed in terms of $\theta$, those results are valid for a given $\theta$ even when it reflects cases with two different masses for the two eikonal lines.

In Section 2, we briefly review results for the cusp anomalous dimension at one, two, and three loops. In Section 3, we discuss the small-$\beta$ expansions of the cusp anomalous dimension through four loops. In Section 4, we discuss the large-$\beta$ behavior of $\Gamma_{\rm cusp}$. In Section 5, we introduce expressions that use the asymptotic behavior at small and large $\beta$, and that numerically describe the cusp anomalous dimension exceedingly well for the full $\beta$ range at two and three loops, and we make a corresponding prediction at four loops. We study in detail the numerical aspects of the expressions through four loops for $n_f=3$, $n_f=4$, and $n_f=5$, and we make brief comments for other $n_f$ values. We also study separate color structures and discuss various extensions of the method. We conclude in Section 6. Appendix A assembles known expressions for the light-like cusp anomalous dimension where color factors and various other constants are also defined, while Appendix B shows the detailed expression for the three-loop massive cusp anomalous dimension.

\mysection{Massive cusp anomalous dimension in QCD at one, two, and three loops}

We begin with a brief overview of results for the massive $\Gamma_{\rm cusp}$ in QCD through three loops. 

\subsection{One loop}

The QCD cusp anomalous dimension at one loop \cite{AMP} is given by
\beq
\Gamma^{(1)}=C_F (\theta \coth\theta -1) \, .
\label{cusp1loop}
\eeq 

This result can be straightforwardly reexpressed in terms of the quark speed $\beta$. 
Noting that $\coth\theta=(1+\beta^2)/(2\beta)$, we define 
\beq
L_{\beta}=\frac{(1+\beta^2)}{2\beta}\ln\left(\frac{1-\beta}{1+\beta}\right) \, .
\label{Lb}
\eeq
Then, the one-loop cusp anomalous dimension written as a function of $\beta$ is given by
\beq
\Gamma^{(1)}=-C_F \left( L_{\beta} +1 \right) \, .
\label{cusp1loopb}
\eeq 

\subsection{Two loops}

Calculations of the QCD cusp anomalous dimension at two loops have a long history. Results for the relevant two-loop diagrams were presented in Ref. \cite{KnSc} in terms of unevaluated double and triple integrals. The two-loop cusp anomalous dimension was calculated in terms of three unevaluated single integrals in Refs. \cite{KRsj,KR1,KR2}, with $n_f$ terms added in Refs. \cite{GPK,KR3}. The result was further refined into one with a single unevaluated integral in Ref. \cite{AGhq}. All these results were given in terms of the cusp angle, $\theta$.

An independent calculation directly in terms of the quark velocity $\beta$ was presented in Ref. \cite{NK2loop}. This calculation provided the first fully analytical result for the two-loop massive cusp anomalous dimension in QCD without any unevaluated integrals. The cusp anomalous dimension at two loops written as a function of $\beta$ is given by \cite{NK2loop,NK2lp,NKtW}
\beqa
\Gamma^{(2)}&=&K_2 \, \Gamma^{(1)}+C_F C_A \left\{\frac{1}{2}+\frac{\zeta_2}{2}
+\frac{1}{2}\ln^2\left(\frac{1-\beta}{1+\beta}\right) \right. 
\nonumber \\ && \hspace{21mm} 
{}+\frac{(1+\beta^2)}{4\beta}\left[\zeta_2\ln\left(\frac{1-\beta}{1+\beta}\right)-\ln^2\left(\frac{1-\beta}{1+\beta}\right) +\frac{1}{3}\ln^3\left(\frac{1-\beta}{1+\beta}\right)
-{\rm Li}_2\left(\frac{4\beta}{(1+\beta)^2}\right)\right] 
\nonumber \\ && \hspace{21mm} 
{}+\frac{(1+\beta^2)^2}{8\beta^2}\left[-\zeta_3-\zeta_2\ln\left(\frac{1-\beta}{1+\beta}\right)-\frac{1}{3}\ln^3\left(\frac{1-\beta}{1+\beta}\right) \right. 
\nonumber \\ && \hspace{45mm} \left. \left.
{}-\ln\left(\frac{1-\beta}{1+\beta}\right) {\rm Li}_2\left(\frac{(1-\beta)^2}{(1+\beta)^2}\right)
+{\rm Li}_3\left(\frac{(1-\beta)^2}{(1+\beta)^2}\right)\right] \right\}\, , 
\label{cusp2loopb}
\eeqa
where $K_2$ \cite{KT82} is given in Eq. (\ref{K2}) of Appendix A.

Furthermore, it was first shown in Ref. \cite{NK2loop} that one can construct an excellent approximation to the complete two-loop result for the cusp anomalous dimension, Eq. (\ref{cusp2loopb}), by using its asymptotic behavior at small and large $\beta$. We note that the method uses the results for $\Gamma_{\rm cusp}$ in terms of $\beta$, and it would not work as well if one used expressions directly in terms of $\theta$ due to the infinite range of the cusp angle, as we will explain in Section 5, although obviously one can later reexpress both the exact and the approximate results in terms of $\theta$. 

The result of Eq. (\ref{cusp2loopb}) for the two-loop cusp anomalous dimension was also rewritten in Ref. \cite{NK2loop} in terms of $\theta$, and is given by  
\beqa
\Gamma^{(2)}&=&K_2 \, \Gamma^{(1)}+C_F C_A \left\{\frac{1}{2}+\frac{\zeta_2}{2}+\frac{\theta^2}{2} 
-\frac{1}{2}\coth\theta\left[\zeta_2\theta+\theta^2
+\frac{\theta^3}{3}+{\rm Li}_2\left(1-e^{-2\theta}\right)\right] \right.
\nonumber \\ && \hspace{31mm} \left.
{}+\frac{1}{2}\coth^2\theta\left[-\zeta_3+\zeta_2\theta+\frac{\theta^3}{3}
+\theta \, {\rm Li}_2\left(e^{-2\theta}\right)
+{\rm Li}_3\left(e^{-2\theta}\right)\right] \right\}\, .
\label{cusp2loop}
\eeqa

\subsection{Three loops}

The QCD cusp anomalous dimension at three loops was calculated in Refs. \cite{GHKM1,GHKM2}. The result was expressed in terms of a number of harmonic polylogarithms of up to weight 5. The result from \cite{GHKM1,GHKM2} was later reexpressed in terms of regular polylogarithms and single integrals of them in Ref. \cite{NK3loopcusp}, and written as
\beq
\Gamma^{(3)}= K_3 \, \Gamma^{(1)}
+2 K_2 \left(\Gamma^{(2)}-K_2 \, \Gamma^{(1)}\right)+C^{(3)} \, ,
\label{cusp3loop}
\eeq
where $K_3$ \cite{MVV04} is given in Eq. (\ref{K3}) of Appendix A,
and $C^{(3)}$ has a long expression which can be found in Eq. (2.13) of Ref. \cite{NK3loopcusp}.

The cusp anomalous dimension at three loops, Eq. (\ref{cusp3loop}), can also written as a function of $\beta$. We have $C^{(3)}=C_F C_A^2 C^{' (3)}$ with $C^{' (3)}$ given explicitly in Eq. (62) of Ref. \cite{NKuni}. We also provide $C^{' (3)}$ in a somewhat improved form in Appendix B.

Furthermore, it was first shown in Ref. \cite{NK3loopcusp} that one can construct an excellent approximation to the complete three-loop result for the cusp anomalous dimension by using its asymptotic behavior at small and large $\beta$, analogously to the two-loop case of \cite{NK2loop}. Again, we note that the method uses the results for $\Gamma_{\rm cusp}$ written in terms of $\beta$, and it would not work as well if one used expressions directly in terms of $\theta$. 

\mysection{Small-$\beta$ expansion of $\Gamma_{\rm cusp}$ through four loops}

For small $\theta$, we can expand the cusp anomalous dimension around $\theta=0$ \cite{KRsj,KR1,KR2,KR3,NK2loop,NK2lp,GHKM1,GHKM2,NK3loopcusp,GLP} as
\beq
\Gamma^{(n)}=\Gamma^{(n)}_{\theta^2}+\Gamma^{(n)}_{\theta^4}+{\cal O}(\theta^6) \, .
\eeq
Expansions at one and two loops were given in \cite{KRsj,KR1,KR2,KR3,NK2loop,NK2lp}, and at three loops in \cite{GHKM1,GHKM2,NK3loopcusp}. The small-$\theta$ expansion at four loops was recently derived in \cite{GLP}.

We note that for small $\theta$, we have $\theta=2\beta+(2/3)\beta^3+{\cal O}(\beta^5)$ and, thus, $\theta^2=4\beta^2+(8/3)\beta^4+{\cal O}(\beta^6)$, so the small $\theta$ expansion formulas can easily be rewritten in terms of $\beta$ \cite{NK2loop,NK3loopcusp}. Equivalently, we have $\beta=\theta/2-\theta^3/24+{\cal O}(\theta^5)$ and, thus, $\beta^2=\theta^2/4-\theta^4/24+{\cal O}(\theta^6)$.

For small $\beta$, we can expand the cusp anomalous dimension around $\beta=0$ \cite{NK2loop,NK2lp,NK3loopcusp} as
\beq
\Gamma^{(n)}=\Gamma^{(n)}_{\beta^2}+\Gamma^{(n)}_{\beta^4}+{\cal O}(\beta^6) \, ,
\eeq
and we find at one loop
\beq
\Gamma^{(1)}_{\beta^2}=\frac{4}{3}C_F \beta^2 \, ,
\label{G1b2}
\eeq
\beq
\Gamma^{(1)}_{\beta^4}=\frac{8}{15}C_F \beta^4 \, ,
\label{G1b4}
\eeq
and at two loops
\beq
\Gamma^{(2)}_{\beta^2}= \beta^2 \left[C_F C_A \left(\frac{94}{27}-\frac{4}{3}\zeta_2 \right)-\frac{20}{27} C_F n_f T_F \right] \, ,
\eeq
\beq
\Gamma^{(2)}_{\beta^4}= \beta^4 \left[C_F C_A \left(\frac{64}{45}-\frac{8}{15}\zeta_2 \right)-\frac{8}{27} C_F n_f T_F \right] \, .
\eeq
We note that if we define $\Gamma^{(1)}_{\beta^{2,4}}=\Gamma^{(1)}_{\beta^2}+\Gamma^{(1)}_{\beta^4}$ and $\Gamma^{(2)}_{\beta^{2,4}}=\Gamma^{(2)}_{\beta^2}+\Gamma^{(2)}_{\beta^4}$, we have the relation
\beq
\Gamma^{(2)}_{\beta^{2,4}}=K_2 \Gamma^{(1)}_{\beta^{2,4}}+\beta^2 C_F C_A \left(1-\frac{2}{3}\zeta_2 \right)+\beta^4 C_F C_A \left(\frac{58}{135}-\frac{4}{15} \zeta_2 \right) \, .
\eeq

At three loops we have
\beqa
\Gamma^{(3)}_{\beta^2}&=& \beta^2 \left[C_F C_A^2 \left(\frac{473}{72}-\frac{170}{27}\zeta_2+\frac{5}{18}\zeta_3+5 \zeta_4\right)
+C_F C_A n_f T_F \left(-\frac{389}{162}+\frac{40}{27} \zeta_2-\frac{14}{9} \zeta_3 \right) \right.
\nonumber \\ && \hspace{10mm} \left.
{}+C_F^2 n_f T_F \left(-\frac{55}{36}+\frac{4}{3} \zeta_3 \right)
-\frac{4}{81} C_F n_f^2 T_F^2 \right] \, ,
\eeqa
\beqa
\Gamma^{(3)}_{\beta^4}&=& \beta^4  \left[C_F C_A^2 \left(\frac{88351}{24300}-\frac{20}{9}\zeta_2-\frac{251}{225}\zeta_3+2 \zeta_4\right)
+C_F C_A n_f T_F \left(-\frac{1207}{1215}+\frac{16}{27} \zeta_2-\frac{28}{45} \zeta_3 \right) \right.
\nonumber \\ && \hspace{10mm} \left.
{}+C_F^2 n_f T_F \left(-\frac{11}{18}+\frac{8}{15} \zeta_3 \right)
-\frac{8}{405} C_F n_f^2 T_F^2 \right] \, .
\eeqa

Using the small-$\theta$ expansion given in Ref. \cite{GLP}, we can derive the small-$\beta$ expansion at four loops, which is given by 
\beqa
\Gamma^{(4)}_{\beta^2}&=& \beta^2 \left[ C_F C_A^3 \left(\frac{89011}{7776}-\frac{17953}{972}\zeta_2+\frac{1189}{324}\zeta_3+\frac{4841}{144}\zeta_4-\frac{155}{72}\zeta_5-\frac{175}{12}\zeta_6-\frac{8}{9} \zeta_2 \zeta_3 \right) \right.
\nonumber \\ && \quad
{}+C_F^2 C_A n_f T_F \left(-\frac{25943}{3888}+\frac{55}{18} \zeta_2 +\frac{170}{27} \zeta_3 -\frac{11}{6}\zeta_4+\frac{5}{3}\zeta_5-\frac{8}{3} \zeta_2 \zeta_3 \right)
\nonumber \\ && \quad
{}+C_F C_A^2 n_f T_F \left(-\frac{48161}{7776}+\frac{1846}{243} \zeta_2 -\frac{3611}{324} \zeta_3 -\frac{55}{9}\zeta_4+\frac{55}{18}\zeta_5+\frac{28}{9} \zeta_2 \zeta_3 \right)
\nonumber \\ && \quad
{}+C_F^3 n_f T_F \left(\frac{143}{216}+\frac{37}{18} \zeta_3 -\frac{10}{3} \zeta_5 \right)
+C_F^2 n_f^2 T_F^2 \left(\frac{299}{486}-\frac{40}{27} \zeta_3 +\frac{2}{3}\zeta_4 \right)
\nonumber \\ && \quad
{}+C_F C_A n_f^2 T_F^2 \left(\frac{1835}{3888}-\frac{76}{243} \zeta_2+\frac{140}{81} \zeta_3 -\frac{7}{9}\zeta_4 \right)
+C_F n_f^3 T_F^3 \left(-\frac{4}{243}+\frac{8}{81} \zeta_3 \right)
\nonumber \\ && \quad \left.
{}+\frac{d_F^{abcd} d_F^{abcd}}{N_c} n_f \left(-\frac{20}{9} \zeta_2 -\frac{50}{3} \zeta_4 +\frac{32}{3} \zeta_2 \zeta_3 \right)
+\frac{d_F^{abcd} d_A^{abcd}}{N_c} \left(-\frac{2}{9} \zeta_2 +\frac{80}{3} \zeta_4 +14 \zeta_6-\frac{68}{3} \zeta_2 \zeta_3 \right) \right]
\nonumber \\
\label{4lb2}
\eeqa 
and
\beqa
\Gamma^{(4)}_{\beta^4}&=& \beta^4 \left[ C_F C_A^3 \left(\frac{42813919}{4374000}-\frac{286153}{36450}\zeta_2-\frac{507971}{60750}\zeta_3+\frac{68987}{5400}\zeta_4+\frac{2351}{540}\zeta_5-\frac{35}{6}\zeta_6+\frac{692}{675} \zeta_2 \zeta_3 \right) \right.
\nonumber \\ && \quad
{}+C_F^2 C_A n_f T_F \left(-\frac{26603}{9720}+\frac{11}{9} \zeta_2 +\frac{116}{45} \zeta_3 -\frac{11}{15}\zeta_4+\frac{2}{3}\zeta_5-\frac{16}{15} \zeta_2 \zeta_3 \right)
\nonumber \\ && \quad
{}+C_F C_A^2 n_f T_F \left(-\frac{17835961}{4374000}+\frac{18821}{6075} \zeta_2 -\frac{16969}{6750} \zeta_3 -\frac{2164}{675}\zeta_4+\frac{181}{135}\zeta_5+\frac{776}{675} \zeta_2 \zeta_3 \right)
\nonumber \\ && \quad
{}+C_F^3 n_f T_F \left(\frac{143}{540}+\frac{37}{45} \zeta_3 -\frac{4}{3} \zeta_5 \right)
+C_F^2 n_f^2 T_F^2 \left(\frac{299}{1215}-\frac{16}{27} \zeta_3 +\frac{4}{15}\zeta_4 \right)
\nonumber \\ && \quad
{}+C_F C_A n_f^2 T_F^2 \left(\frac{17123}{87480}-\frac{152}{1215} \zeta_2+\frac{56}{81} \zeta_3 -\frac{14}{45}\zeta_4 \right)
+C_F n_f^3 T_F^3 \left(-\frac{8}{1215}+\frac{16}{405} \zeta_3 \right)
\nonumber \\ && \quad
{}+\frac{d_F^{abcd} d_F^{abcd}}{N_c} n_f \left(-\frac{92}{225}-\frac{752}{225} \zeta_2 +\frac{1136}{225} \zeta_3 -\frac{12}{5} \zeta_4-\frac{64}{9} \zeta_5+\frac{1088}{225} \zeta_2 \zeta_3 \right)
\nonumber \\ && \quad \left.
{}+\frac{d_F^{abcd} d_A^{abcd}}{N_c} \left(\frac{32}{243}-\frac{6892}{1215} \zeta_2 +\frac{2264}{405} \zeta_3+\frac{56}{45} \zeta_4 -\frac{56}{9} \zeta_5 +\frac{28}{5} \zeta_6+\frac{104}{225} \zeta_2 \zeta_3 \right) \right] \, .
\nonumber \\
\label{4lb4}
\eeqa 

\mysection{Large-$\beta$ behavior of $\Gamma_{\rm cusp}$}

The massless limit, $m \to 0$, of the cusp anomalous dimension, which is
the limit $\theta \to \infty$, is given in Eq. (\ref{lightlike}). Equivalently, this is the limit $\beta \to 1$, and it can be written as
\beq
\lim_{\beta \to 1} \Gamma^{(n)}= K_n  \lim_{\beta \to 1} \Gamma^{(1)}+ P_n \, ,
\label{masslessP}
\eeq
where $K_n$ for $n=1,2,3,4$ are given in Appendix A, and the constants $P_n$ at one, two, and three loops are given, respectively, by 
$P_1=0$, $P_2=(1/2) C_F C_A (1-\zeta_3)$,
and 
\beq
P_3=K_2 C_F C_A (1-\zeta_3)
+C_F C_A^2 \left(-\frac{1}{2}+\frac{3}{4}\zeta_2-\frac{\zeta_3}{4}
+\frac{9}{8} \zeta_5-\frac{3}{4}\zeta_2 \zeta_3 \right) \, .
\label{P3}
\eeq

The limit can also be rewritten as
\beq
\lim_{\beta \to 1} \Gamma^{(n)}= - C_F K_n  \lim_{\beta \to 1} \ln\left(\frac{1-\beta}{2}\right) + R_n 
= - C_F K_n \lim_{m \to 0} \ln\left(\frac{m^2}{s}\right) + R_n            \, ,
\label{masslessR}
\eeq
where the constants $R_n$ are given by $R_n=P_n-C_F K_n$.

\mysection{Expressions for $\Gamma_{\rm cusp}$ through four loops from asymptotics}

As first shown in Ref. \cite{NK2loop} for the two-loop case, we can construct simple expressions based on the asymptotics of $\Gamma_{\rm cusp}$ that provide excellent approximations which are valid for all values of $\beta$. At all orders, the cusp anomalous dimension vanishes at $\beta=0$ and is infinite at $\beta=1$.
The expansion around $\beta=0$ gives very good approximations to $\Gamma^{(n)}$ at small $\beta$. 
The expression in Eq. (\ref{masslessP}) gives the large $\beta$ limit, which 
shows that in that limit the higher-loop results are essentially proportional to the one-loop result. Thus, we can derive an approximate expression from asymptotics, denoted as $\Gamma^{(n)}_A$, for all $\beta$ values by starting with the small $\beta$ expansion of $\Gamma^{(n)}$, then adding $K_n \, \Gamma^{(1)}$ and subtracting from it its small $\beta$ expansion:
\beq
\Gamma^{(n)}_A=\Gamma^{(n)}_{\beta^{2,4}}-K_n \, \Gamma^{(1)}_{\beta^{2,4}}
+K_n \, \Gamma^{(1)} 
\label{GammaA}
\eeq
where $\Gamma^{(n)}_{\beta^{2,4}}=\Gamma^{(n)}_{\beta^2}+\Gamma^{(n)}_{\beta^4}$. We note that the last two terms on the right in the above equation cancel precisely against each other at small $\beta$, and quite well even at medium $\beta$, while the first  
two terms largely cancel against each other at large $\beta$.

Equivalently, using Eqs. (\ref{cusp1loopb}), (\ref{G1b2}), and (\ref{G1b4}), we can write Eq. (\ref{GammaA}) as
\beq
\Gamma^{(n)}_A=\Gamma^{(n)}_{\beta^{2,4}}
-C_F K_n \left(\frac{4}{3}\beta^2+\frac{8}{15}\beta^4+L_{\beta}+1 \right) \, .
\label{GammaAp}
\eeq

We note that for the one-loop case, we have $\Gamma^{(1)}_A=\Gamma^{(1)}$ identically. Applying Eq. (\ref{GammaA}) to higher loops, setting the number of colors $N_c=3$, and numerically evaluating all constants, we find very simple expressions in terms of $\beta$ and $n_f$ at two, three, and four loops:
\beq
\Gamma^{(2)}_A = -0.386490845 \, \beta^2 - 0.036077819 \, \beta^4  + (3.115932233 - 0.277777778 \, n_f) \; \Gamma^{(1)} \, ,
\label{GammaA2}
\eeq
\beqa
\Gamma^{(3)}_A&=& (-0.981370903 + 0.214717136 \, n_f) \, \beta^2 + (-0.141381392 + 0.020043233 \, n_f) \, \beta^4 
\nonumber \\ &&
{}+ (13.76833912 - 2.146727700 \, n_f - 0.009259259 \, n_f^2) \; \Gamma^{(1)} \, ,
\label{GammaA3}
\eeqa
\beqa
\Gamma^{(4)}_A&=& (-3.749290323 + 1.186688634 \, n_f - 0.022664587 \, n_f^2) \, \beta^2 
\nonumber \\ &&
{}+(-0.290594150 + 0.156331101 \, n_f - 0.002115675 \, n_f^2) \, \beta^4
\nonumber \\ &&
{}+ (60.65142489 - 15.15209803 \, n_f + 0.572980154 \, n_f^2 + 0.009586947 \, n_f^3) \; \Gamma^{(1)} \, ,
\label{GammaA4}
\eeqa
where $\Gamma^{(1)}$ is given by Eq. (\ref{cusp1loopb}) with $C_F=4/3$ in QCD.

We note that the $n_f$ terms in $\Gamma^{(2)}_A$ are 
$\Gamma^{(2) \, n_f}_A=-(5/9) n_f T_F \Gamma^{(1)}$, so they are identically the same as in the exact result, but the $C_F C_A$ terms are not exact. We also note that the $n_f$ terms in $\Gamma^{(3)}_A$ are $\Gamma^{(3) \, n_f}_A=K_3^{n_f} \Gamma^{(1)}+2 K_2^{n_f} \, \left(\Gamma^{(2)}_{\beta^{2,4}}-K_2 \, \Gamma^{(1)}_{\beta^{2,4}}\right)$, where $K_2^{n_f}$ and $K_3^{n_f}$ denote the $n_f$ terms in $K_2$ and $K_3$. Thus, in $\Gamma^{(3)}_A$  the $C_F^2 n_f$ and the $C_F n_f^2$  terms are exact but the $C_F C_A^2$ and the $C_F C_A n_f$ terms are not exact. Finally, at four loops, the $C_F^3 n_f$, $C_F^2 n_f^2$, and $C_F n_f^3$ terms in  $\Gamma^{(4)}_A$ are exact, but all the rest of the terms are not exact.

As mentioned earlier, the method would not work well directly in terms of $\theta$, i.e. if the above expressions used $\theta^2$ and $\theta^4$ expansions and $\Gamma^{(1)}$ in terms of $\theta$; this is due to the infinite range of the cusp angle which would result in incomplete cancellations and poor results at large $\theta$. Thus, the method has to be used exactly as described above, which benefits from the finite and small $\beta$ range of 0 to 1. Of course, at the end one can still reexpress Eqs. (\ref{GammaA}) through (\ref{GammaA4})  in terms of $\theta$ with the simple substitution $\beta=\tanh(\theta/2)$. 

\subsection{Results for $n_f=3$}

We begin our numerical study of the cusp anomalous dimension through four loops for the case $n_f=3$, i.e. three light-quark flavors. This would, for example, be relevant to charm pair production via $e^+ e^- \to c \, {\bar c}$.

\begin{figure}[t]
\begin{center}
\includegraphics[width=135mm]{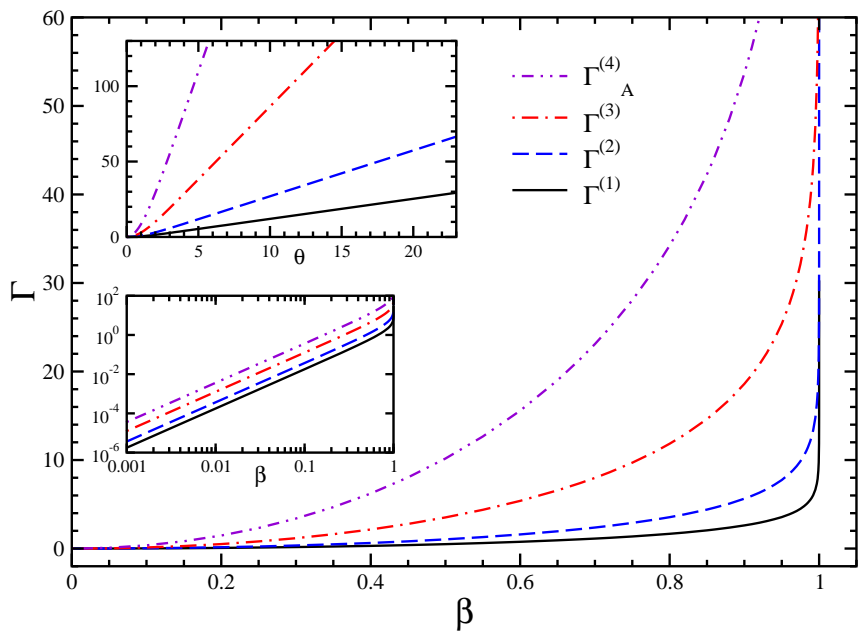}
\vspace{-5mm}
\caption{The cusp anomalous dimension for $n_f=3$.}
\label{cuspnf3}
\end{center}
\end{figure}

In Fig. \ref{cuspnf3} we plot the cusp anomalous dimension for $n_f=3$ at one, two, three, and four loops as a function of $\beta$. The one-loop $\Gamma^{(1)}$, two-loop $\Gamma^{(2)}$, and three-loop $\Gamma^{(3)}$ results are exact, while the four-loop result $\Gamma^{(4)}_A$ is the expression from the asymptotics in Eqs. (\ref{GammaA}) and (\ref{GammaA4}). To better show the behavior for small $\beta$, we plot the results in a logarithmic scale over several orders of magnitude in the lower inset plot. On the other hand, to better show the behavior near $\beta=1$, we plot the results as functions of the cusp angle $\theta$ in the upper inset plot. For example, a value of $\beta=0.99999$ corresponds to $\theta \approx 12.2$. Thus, the three different ways of plotting the results give an overall picture of the behavior of $\Gamma_{\rm cusp}$ for small, medium, and large $\beta$ values through four loops.

\begin{figure}[t]
\begin{center}
\includegraphics[width=135mm]{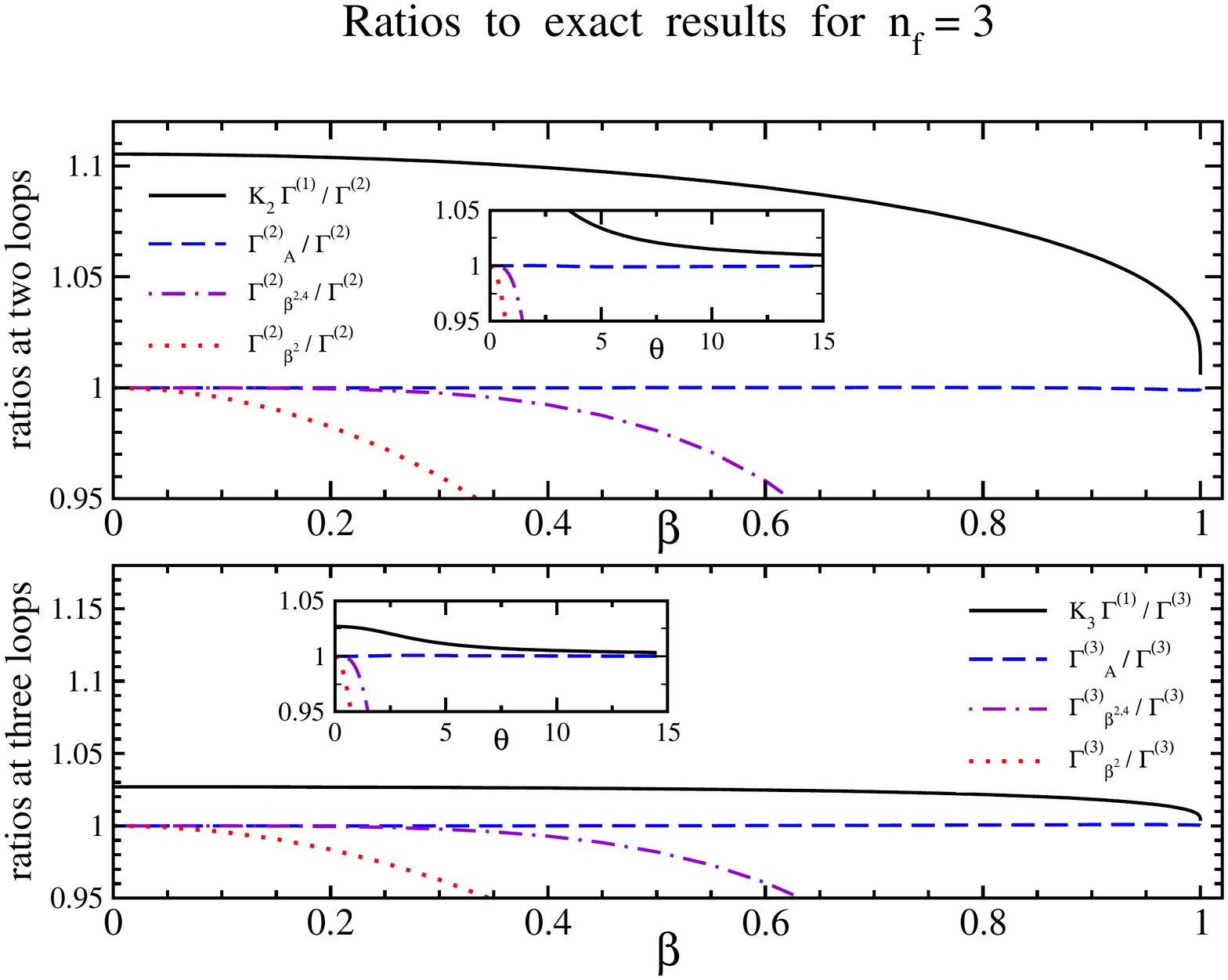}
\vspace{-5mm}
\caption{Ratios for the cusp anomalous dimension with $n_f=3$ at two loops (upper plot) and three loops (lower plot).}
\label{Kcuspnf3}
\end{center}
\end{figure}

In Fig. \ref{Kcuspnf3} we plot ratios of the various terms in Eq. (\ref{GammaA}) to the exact result for the cusp anomalous dimension at two and three loops for $n_f=3$. The upper plot of Fig. \ref{Kcuspnf3} shows ratios at two loops. The ratio $K_2 \, \Gamma^{(1)}/\Gamma^{(2)}$ approaches the value 1 at large $\beta$, as expected, but it is considerably larger than that for most of the $\beta$ range, so by itself it is not an adequate approximation of the exact two-loop result. The small-$\beta$ approximation is a good approximation at small $\beta$, as expected, but it begins to fail at larger values. The $\Gamma^{(2)}_{\beta^2}/\Gamma^{(2)}$ ratio shows that the $\beta^2$ terms by themselves provide a description of the exact result by better than one part in ten thousand (i.e. 0.1 per mille) up to $\beta \approx 0.015$, and better than one per mille up to $\beta \approx 0.05$. The $\Gamma^{(2)}_{\beta^{2,4}}/\Gamma^{(2)}$ ratio shows that the sum of the $\beta^2$ and $\beta^4$ terms provides a description better than 0.1 per mille up to $\beta \approx 0.14$, and better than one per mille up to $\beta \approx 0.24$. The expansions begin to fail at higher values of $\beta$. By a value of $\beta \approx 0.6$, even $\Gamma^{(2)}_{\beta^{2,4}}$ differs by four percent from $\Gamma^{(2)}$. The result for $\Gamma^{(2)}_A$, however, provides an excellent description throughout the $\beta$ range, as the ratio $\Gamma^{(2)}_A/\Gamma^{(2)}$ shows. The difference between $\Gamma^{(2)}_A$ and $\Gamma^{(2)}$ is less than one per mille over the entire $\beta$ range from 0 to 1; in fact, it is less than one part per million from $\beta=0$ up to $\beta \approx$ 0.17, and better than 0.1 per mille for most of the $\beta$ range, from $\beta=0$ to $\beta \approx 0.6$, and also for values between $\beta \approx 0.8$ and $\beta \approx 0.9$, as well as for $\beta$ values extremely close to 1. The inset of the upper plot of Fig. \ref{Kcuspnf3} shows the same two-loop ratios as functions of $\theta$ for $n_f=3$. Thus, we see that $\Gamma_A^{(2)}$ performs exceptionally well, by any reasonable standard, in giving the correct prediction for the two-loop cusp anomalous dimension for all $\beta$ values or, equivalently, for all $\theta$ values. The line $\Gamma_A^{(2)}/\Gamma^{2)}$ is practically indistinguishable from 1 in the plots. 

The lower plot of Fig. \ref{Kcuspnf3} shows ratios at three loops for $n_f=3$. The ratio $K_3 \, \Gamma^{(1)}/\Gamma^{(3)}$ approaches the value 1 at large $\beta$, as expected, and it actually remains within three percent of the exact result over the entire $\beta$ range. As also expected, the small-$\beta$ approximation is a good approximation at small $\beta$ but not at larger values. The $\Gamma^{(3)}_{\beta^2}/\Gamma^{(3)}$ ratio shows that the $\beta^2$ terms by themselves provide a description of better than 0.1 per mille up to $\beta \approx 0.016$, and better than one per mille up to $\beta \approx 0.05$, which is very similar to what we saw at two loops above. The $\Gamma^{(3)}_{\beta^{2,4}}/\Gamma^{(3)}$ ratio shows that the sum of the $\beta^2$ and $\beta^4$ terms provides a description of better than 0.1 per mille up to $\beta \approx 0.14$, and better than one per mille up to $\beta \approx 0.25$, which again is very similar to the behavior at two loops. By a value of $\beta \approx 0.6$, however, $\Gamma^{(3)}_{\beta^{2,4}}$ differs by four percent from $\Gamma^{(3)}$. On the other hand, as the ratio $\Gamma^{(3)}_A/\Gamma^{(3)}$ shows, $\Gamma^{(3)}_A$ provides an excellent description over the entire $\beta$ range. The difference between $\Gamma^{(3)}_A$ and $\Gamma^{(3)}$ stays well below one per mille everywhere; in fact, it is less than one part per million from $\beta=0$ up to $\beta \approx 0.16$, and better than 0.1 per mille for the majority of the $\beta$ range, from $\beta=0$ to above $\beta \approx 0.5$ as well as for $\beta$ values extremely close to 1. The inset of the lower plot of Fig. \ref{Kcuspnf3} shows the same three-loop ratios as functions of $\theta$ for $n_f=3$. Thus, we see that $\Gamma_A^{(3)}$ performs exceptionally well in giving the correct prediction for the three-loop cusp anomalous dimension over all $\beta$ or $\theta$ values. The line $\Gamma_A^{(3)}/\Gamma^{3)}$ is virtually indistinguishable from 1 in the plots. 

The great similarity between the two-loop and three-loop cases in the behavior of the expansions with $\beta^2$ and $\beta^4$ terms and, more importantly, of the approximate expressions from asymptotics (despite the difference in the ratios $K_2 \, \Gamma^{(1)}/\Gamma^{(2)}$ and $K_3 \, \Gamma^{(1)}/\Gamma^{(3)}$), indicates a very strong robustness of our method for calculating $\Gamma^{(n)}_A$. The fact that $\Gamma^{(2)}_A$ and $\Gamma^{(3)}_A$ are practically indistinguishable from the corresponding exact results highlights the success of the formula in Eq. (\ref{GammaA}) and gives strong confidence for its success at higher loops. 

\begin{figure}[t]
\begin{center}
\includegraphics[width=135mm]{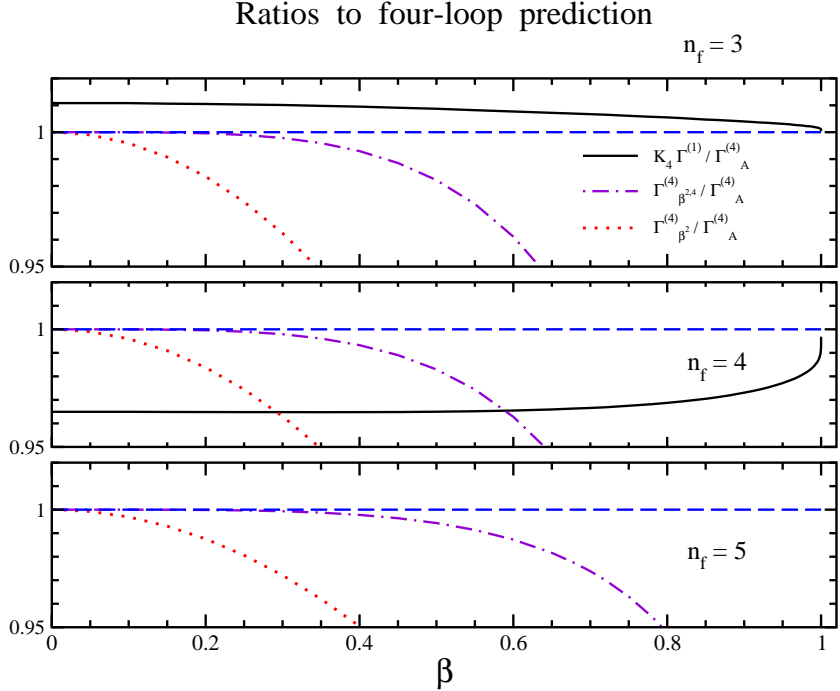}
\vspace{-5mm}
\caption{Ratios with respect to $\Gamma^{(4)}_A$ for $n_f=3$ (top plot), $n_f=4$ (middle plot), and $n_f=5$ (lower plot).}
\label{K4cusp}
\end{center}
\end{figure}

Since we do not know the full exact result for $\Gamma^{(4)}$, we cannot create an exact analog of Fig. \ref{Kcuspnf3} at four loops. However, we can do something similar and study the ratio $K_4 \, \Gamma^{(1)}/\Gamma^{(4)}_A$ as well as the ratios of the small-$\beta$ expansions to $\Gamma^{(4)}_A$. In the top plot of Fig. \ref{K4cusp} we plot these ratios for $n_f=3$. We also plot the dashed line identically equal to 1 for reference, and note that we expect it to be practically indistinguishable from the ratio $\Gamma^{(4)}_A/\Gamma^{(4)}$. The $K_4 \, \Gamma^{(1)}$ term by itself is very close to $\Gamma^{(4)}_A$, around one percent or better over the entire range. We observe that the behavior of the small-$\beta$ expansions is very similar to the two-loop and three-loop cases, again displaying consistency across different orders. The $\Gamma^{(4)}_{\beta^2}/\Gamma^{(4)}_A$ ratio shows that the $\beta^2$ terms provide a description better than 0.1 per mille up to $\beta \approx 0.016$, and better than one per mille up to $\beta \approx 0.05$, which is very similar to what we saw at two and three loops above. The $\Gamma^{(4)}_{\beta^{2,4}}/\Gamma^{(4)}_A$ ratio shows that the sum of the $\beta^2$ and $\beta^4$ terms provides a description better than 0.1 per mille up to $\beta \approx 0.14$, and better than one per mille up to $\beta \approx 0.25$, which again is very similar to the behavior at two loops and at three loops. By a value of $\beta \approx 0.6$, $\Gamma^{(4)}_{\beta^{2,4}}$ differs by four percent from $\Gamma^{(4)}_A$. Again, all this behavior is very similar to the situation at two and three loops, and it highlights the robustness of the approach and provides strong confidence that the result for $\Gamma^{(4)}_A$ is numerically essentially the same as that for $\Gamma^{(4)}$ for all practical purposes.

\subsection{Results for $n_f=4$}

We continue our numerical study of the cusp anomalous dimension through four loops for the case $n_f=4$, i.e. four light-quark flavors. This would, for example, be relevant to $b$-quark pair production via $e^+ e^- \to b \, {\bar b}$.

\begin{figure}[t]
\begin{center}
\includegraphics[width=135mm]{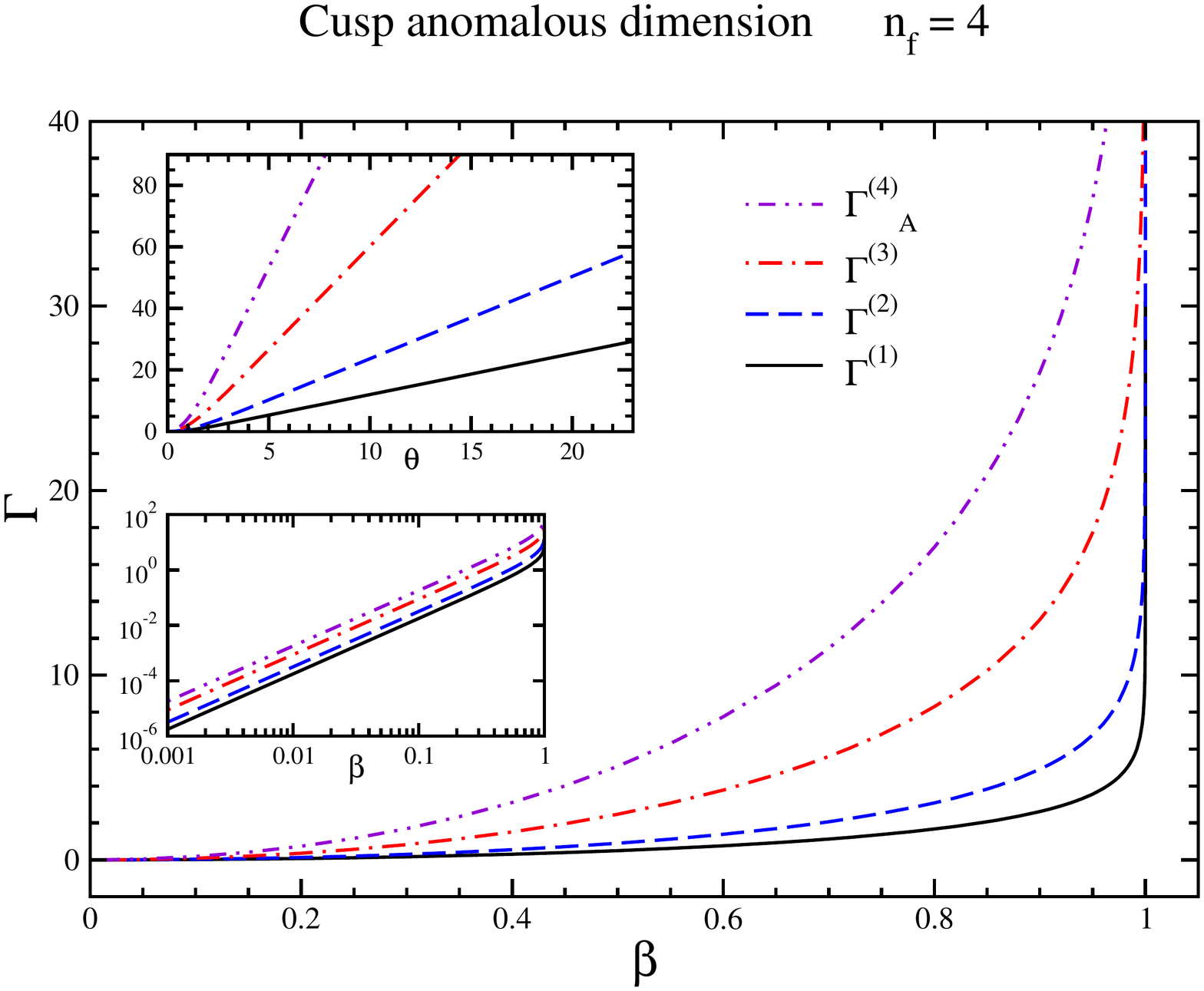}
\vspace{-5mm}
\caption{The cusp anomalous dimension for $n_f=4$.}
\label{cuspnf4}
\end{center}
\end{figure}

In Fig. \ref{cuspnf4} we plot the cusp anomalous dimension for $n_f=4$ as a function of $\beta$.  As before, the one-loop $\Gamma^{(1)}$, two-loop $\Gamma^{(2)}$, and three-loop $\Gamma^{(3)}$ results are exact, while the four-loop result $\Gamma^{(4)}_A$ is the expression from the asymptotics. Of course, since the one-loop result is independent of $n_f$, it is identical to what we already plotted in Fig. \ref{cuspnf3}, but for higher loops the results differ and, thus, the vertical scales used in the plots of Fig. \ref{cuspnf4} are different from those in Fig. \ref{cuspnf3}. Again, to better show the behavior for small $\beta$, we plot the results in a logarithmic scale over several orders of magnitude in the lower inset plot, while to better show the behavior near $\beta=1$, we plot the results as functions of the cusp angle $\theta$ in the upper inset plot. 

\begin{figure}[t]
\begin{center}
\includegraphics[width=135mm]{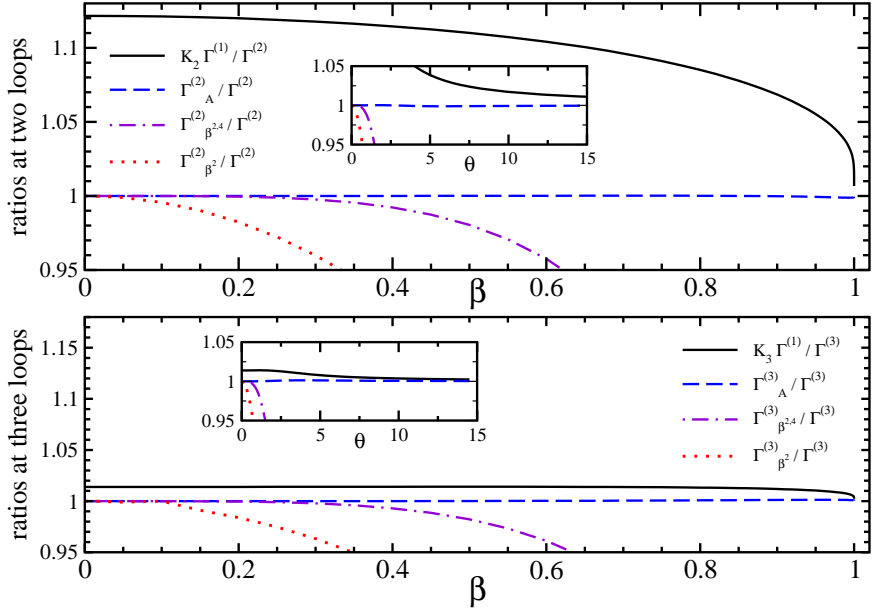}
\vspace{-5mm}
\caption{Ratios for the cusp anomalous dimension with $n_f=4$ at two loops (upper plot) and three loops (lower plot).}
\label{Kcuspnf4}
\end{center}
\end{figure}

In Fig. \ref{Kcuspnf4} we plot ratios of the various terms in Eq. (\ref{GammaA}) to the exact result for the cusp anomalous dimension at two and three loops for $n_f=4$. The upper plot of Fig. \ref{Kcuspnf4} shows ratios at two loops. The ratio $K_2 \, \Gamma^{(1)}/\Gamma^{(2)}$ approaches the value 1 at large $\beta$, as expected, but it is considerably larger than that for most of the $\beta$ range. $\Gamma^{(2)}_{\beta^2}$ differs from $\Gamma^{(2)}$ by less than 0.1 per mille up to $\beta \approx 0.015$, and less than one per mille up to $\beta \approx 0.05$. $\Gamma^{(2)}_{\beta^{2,4}}$ differs from $\Gamma^{(2)}$ by less than 0.1 per mille up to $\beta \approx 0.14$, and less than one per mille up to $\beta \approx 0.24$. By a value of $\beta \approx 0.6$, $\Gamma^{(2)}_{\beta^{2,4}}$ differs by more than four percent from $\Gamma^{(2)}$. This is all very similar to the small-$\beta$ asymptotic behavior for $n_f=3$ as we saw in the previous subsection. Moreover, the result for $\Gamma^{(2)}_A$ provides an excellent description throughout the $\beta$ range as the ratio $\Gamma^{(2)}_A/\Gamma^{(2)}$ shows. The difference between $\Gamma^{(2)}_A$ and $\Gamma^{(2)}$ is one per mille or less over the entire $\beta$ range from 0 to 1; in fact, it is less than one part per million from $\beta=0$ up to $\beta \approx$ 0.16, and 0.1 per mille or better for most of the $\beta$ range, from $\beta=0$ to $\beta \approx 0.6$, and also for values between $\beta \approx 0.8$ and $\beta \approx 0.9$, as well as for $\beta$ values extremely close to 1. Again, these results are very similar to the corresponding ones for $n_f=3$. The inset of the upper plot of Fig. \ref{Kcuspnf4} shows the same two-loop ratios as functions of $\theta$ for $n_f=4$. 

The lower plot of Fig. \ref{Kcuspnf4} shows ratios at three loops for $n_f=4$. The ratio $K_3 \, \Gamma^{(1)}/\Gamma^{(3)}$ approaches the value 1 at large $\beta$, as expected, and it actually remains within one-and-a-half percent of the exact result over the entire $\beta$ range. The $\Gamma^{(3)}_{\beta^2}$ terms differ from $\Gamma^{(3)}$ by less than 0.1 per mille up to $\beta\approx 0.016$, and less than one per mille up to $\beta \approx 0.05$, which is very similar to what we saw at two loops. The $\Gamma^{(3)}_{\beta^{2,4}}$ terms differ from $\Gamma^{(3)}$ by less than 0.1 per mille up to $\beta \approx 0.14$, and less than one per mille up to $\beta \approx 0.25$, which again is very similar to the behavior at two loops. The result for $\Gamma^{(3)}_A$ provides an excellent description over the entire $\beta$ range. The difference between $\Gamma^{(3)}_A$ and $\Gamma^{(3)}$ is one per mille or better everywhere; it is actually less than one part per million from $\beta=0$ up to $\beta \approx$ 0.15, and 0.1 per mille or better for half of the $\beta$ range, from $\beta=0$ to $\beta \approx 0.5$ as well as for $\beta$ values extremely close to 1. Again, this is very similar to what we saw for the $n_f=3$ case. The inset of the lower plot of Fig. \ref{Kcuspnf4} shows the same three-loop ratios as functions of $\theta$. Thus, $\Gamma_A^{(3)}$ again performs exceptionally well in giving the correct prediction for the three-loop cusp anomalous dimension for $n_f=4$.

Again, since we do not know the full exact result for $\Gamma^{(4)}$, we cannot create a direct analog of Fig. \ref{Kcuspnf4} at four loops. However, we can study the ratio $K_4 \, \Gamma^{(1)}/\Gamma^{(4)}_A$ as well as the ratios of the small-$\beta$ expansions to $\Gamma^{(4)}_A$. In the middle plot of Fig. \ref{K4cusp} we plot these ratios for $n_f=4$, with the dashed line identically equal to 1 for reference. The $K_4 \, \Gamma^{(1)}$ term by itself is somewhat smaller than $\Gamma^{(4)}_A$. Also, $\Gamma^{(4)}_{\beta^2}$ differs from $\Gamma^{(4)}_A$ by less than 0.1 per mille up to $\beta \approx 0.016$, and less than one per mille up to $\beta \approx 0.05$, which is very similar to what we saw at two and three loops. $\Gamma^{(4)}_{\beta^{2,4}}$ differs from $\Gamma^{(4)}_A$ by less than 0.1 per mille up to $\beta \approx 0.14$, and less than one per mille up to $\beta \approx 0.25$. Once again, all this behavior is very similar to the situation at two and three loops for both $n_f=3$ and $n_f=4$, as well as the four-loop results for $n_f=3$, and it provides strong confidence that the result for $\Gamma^{(4)}_A$ is numerically essentially the same as that for $\Gamma^{(4)}$ also for $n_f=4$ for all practical purposes.

\subsection{Results for $n_f=5$}

We continue our numerical study of the cusp anomalous dimension through four loops for the case $n_f=5$, i.e. five light-quark flavors. This would, for example, be relevant to top-quark pair production via $e^+ e^- \to t \, {\bar t}$.

\begin{figure}[t]
\begin{center}
\includegraphics[width=135mm]{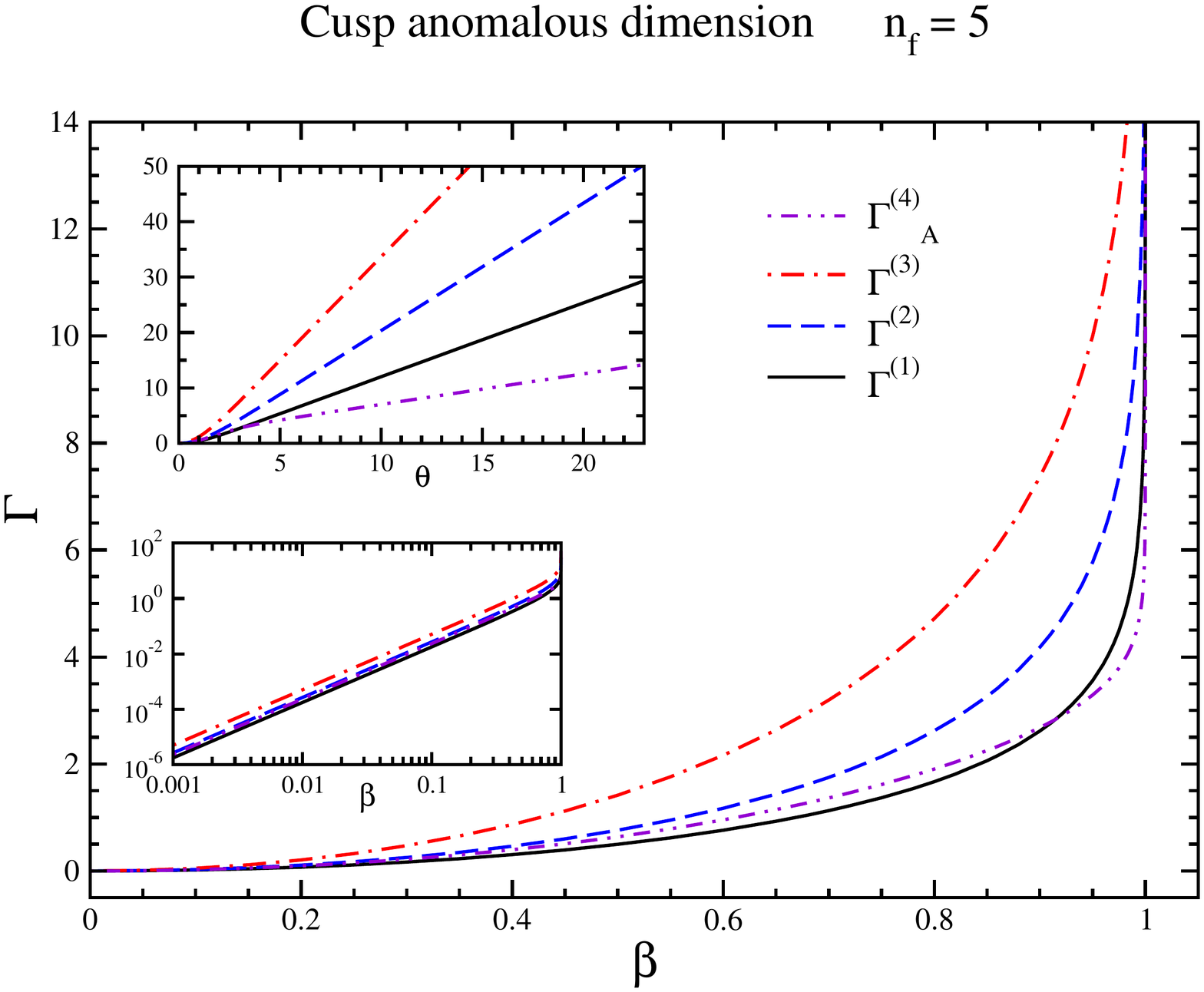}
\vspace{-5mm}
\caption{The cusp anomalous dimension for $n_f=5$.}
\label{cuspnf5}
\end{center}
\end{figure}

In Fig. \ref{cuspnf5} we plot the cusp anomalous dimension for $n_f=5$ as a function of $\beta$. As in the previous cases of Figs. \ref{cuspnf3} and \ref{cuspnf4}, the one-loop $\Gamma^{(1)}$, two-loop $\Gamma^{(2)}$, and three-loop $\Gamma^{(3)}$ results are exact, while the four-loop result $\Gamma^{(4)}_A$ is the expression from the asymptotics. As we have discussed, the one-loop result is the same as before, but for higher loops the results differ, and the vertical scales used in the plots of Fig. \ref{cuspnf5} are different from those in the other cases. The lower inset plot shows more clearly the small-$\beta$ asymptotics in a logarithmic scale, while the upper inset plot shows the results versus $\theta$ in order to show more clearly the behavior near $\beta=1$. 

\begin{figure}[t]
\begin{center}
\includegraphics[width=135mm]{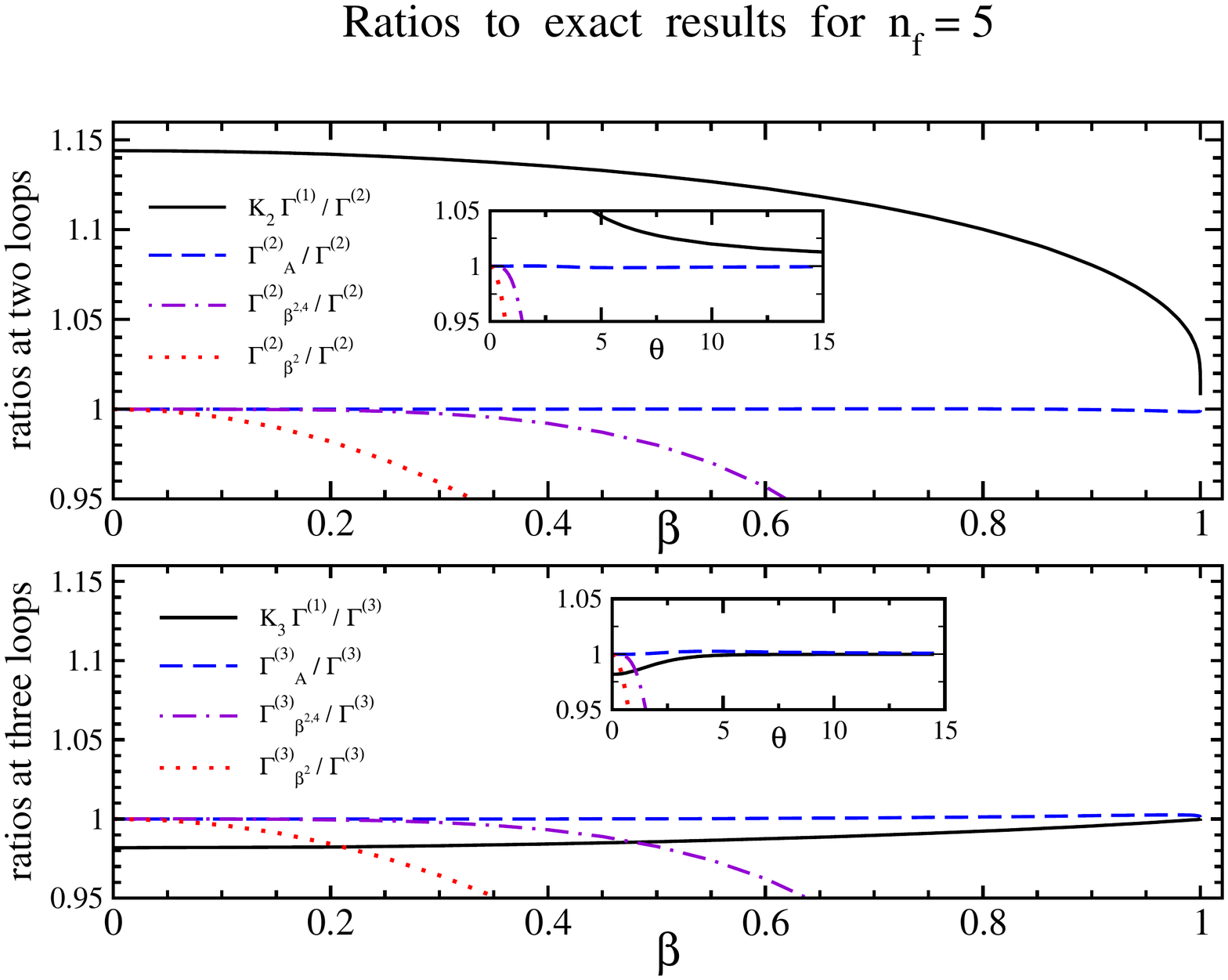}
\vspace{-5mm}
\caption{Ratios for the cusp anomalous dimension with $n_f=5$ at two loops (upper plot) and three loops (lower plot).}
\label{Kcuspnf5}
\end{center}
\end{figure}

In Fig. \ref{Kcuspnf5} we plot ratios of the various terms in Eq. (\ref{GammaA}) to the exact result for the cusp anomalous dimension at two and three loops for $n_f=5$. As before, the upper plot of Fig. \ref{Kcuspnf5} shows ratios at two loops. The ratio $K_2 \, \Gamma^{(1)}/\Gamma^{(2)}$ is considerably larger than 1 for most of the $\beta$ range but tends to 1 at large $\beta$. The $\Gamma^{(2)}_{\beta^2}$ terms differ from $\Gamma^{(2)}$ by less than 0.1 per mille up to $\beta \approx 0.015$, and less than one per mille up to $\beta \approx 0.05$. The $\Gamma^{(2)}_{\beta^{2,4}}$ terms differ from $\Gamma^{(2)}$ by less than 0.1 per mille up to $\beta \approx 0.14$, and less than one per mille up to $\beta \approx 0.24$. Also, the result for $\Gamma^{(2)}_A$ provides an excellent description throughout the $\beta$ range. The difference between $\Gamma^{(2)}_A$ and $\Gamma^{(2)}$ is one per mille or less over the entire $\beta$ range from 0 to 1; indeed, it is less than one part per million from $\beta=0$ up to $\beta \approx 0.16$, and better than 0.1 per mille for most of the $\beta$ range, from $\beta=0$ to above $\beta \approx 0.5$, and also for values between $\beta \approx 0.8$ and $\beta \approx 0.9$, as well as for $\beta$ values extremely close to 1. These results are very similar to the corresponding ones for $n_f=3$ and $n_f=4$, again highlighting the robustness and success of the method. The inset of the upper plot of Fig. \ref{Kcuspnf5} shows the same two-loop ratios as functions of $\theta$ for $n_f=5$. 

The lower plot of Fig. \ref{Kcuspnf5} shows ratios at three loops for $n_f=5$. The ratio $K_3 \, \Gamma^{(1)}/\Gamma^{(3)}$ approaches the value 1 at large $\beta$, as expected, and it remains within two percent of the exact result over the entire $\beta$ range. $\Gamma^{(3)}_{\beta^2}$ differs from $\Gamma^{(3)}$  by less than 0.1 per mille up to $\beta \approx 0.016$, and less than one per mille up to $\beta \approx 0.05$, which is very similar to what we saw at two loops. $\Gamma^{(3)}_{\beta^{2,4}}$ differs from $\Gamma^{(3)}$ by less than 0.1 per mille up to $\beta \approx 0.14$, and less than one per mille up to $\beta \approx 0.25$, which is also very similar to the behavior at two loops. $\Gamma^{(3)}_A$ provides an excellent description over the entire $\beta$ range. The difference between $\Gamma^{(3)}_A$ and $\Gamma^{(3)}$ is well below three per mille everywhere; it is actually less than one part per million from $\beta=0$ up to $\beta \approx$ 0.14, and 0.1 per mille or better for half of the $\beta$ range, from $\beta=0$ to $\beta \approx 0.5$ as well as for $\beta$ values extremely close to 1. Again, this is similar to what we observed in the $n_f=3$ and $n_f=4$ cases. The inset of the lower plot of Fig. \ref{Kcuspnf5} shows the same three-loop ratios as functions of $\theta$. Thus, we observe that $\Gamma_A^{(3)}$ performs exceptionally well in giving the correct prediction for the three-loop cusp anomalous dimension for $n_f=5$.

As discussed before, since we do not know the full exact result for $\Gamma^{(4)}$, we do not have a direct analog of Fig. \ref{Kcuspnf5} at four loops. In the bottom plot of Fig. \ref{K4cusp} we plot ratios at four loops for $n_f=5$, with the dashed line identically equal to 1 for reference. The $\Gamma^{(4)}_{\beta^2}$ terms differ from $\Gamma^{(4)}_A$ by less than 0.1 per mille to $\beta \approx 0.018$, and less than one per mille up to $\beta \approx 0.06$. The $\Gamma^{(4)}_{\beta^{2,4}}$ terms differ from $\Gamma^{(4)}_A$ by less than 0.1 per mille up to $\beta \approx 0.18$, and less than one per mille up to $\beta \approx 0.32$. All this is again similar to the previous cases, and it provides strong confidence in the result for $\Gamma^{(4)}_A$ for $n_f=5$.

\subsection{Results for other values of $n_f$}

Finally, we consider other values for $n_f$, even ones not realized in nature but possibly used in toy models or in models of physics beyond the Standard Model. In fact, we have calculated the cusp anomalous dimension for integer values of $n_f$ ranging from 0 to 10. The results are remarkably consistent in that Eq. (\ref{GammaA}) always provides an excellent approximation to the exact results at two and three loops, throughout the $\beta$ range, and we derive robust and precise four-loop predictions for the cusp anomalous dimension from its asymptotics via Eq. (\ref{GammaA}). 

\subsection{Extensions of the expressions and method}

The method presented in this paper can be extended in a number of ways. One obvious extension is to include more (or fewer) terms in the small-$\beta$ expansion contribution to Eq. (\ref{GammaA}). We can write that relation more generally as 
\beq
\Gamma^{(n)}_A=\Gamma^{(n)}_{{\rm small}-\beta}-K_n \, \Gamma^{(1)}_{{\rm small}-\beta}
+K_n \, \Gamma^{(1)} 
\label{GammaAext}
\eeq
where we can keep as many terms in the small-$\beta$ expansion as we wish.

For example, in Ref. \cite{NK2loop} results were presented using Eq. (\ref{GammaAext}) for $n_f=5$ at two loops with a couple of different choices. On one hand, results were given with only $\beta^2$ terms included in Eq. (\ref{GammaAext}). As shown in \cite{NK2loop}, this is still a good approximation over all $\beta$ values, only about half of one percent or better from the exact value. On the other hand, results were also given in \cite{NK2loop} with terms included through $\beta^{12}$ which of course provide a better approximation. However, there is an issue of diminishing returns. While the inclusion of both $\beta^2$ and $\beta^4$ terms provides small but significant improvements relative to only $\beta^2$ terms in the numerical result from Eq. (\ref{GammaAext}), further additional terms provide negligible impact while affecting the simplicity of our approach.

We also note that in Ref. \cite{NK3loopcusp} results were presented using Eq. (\ref{GammaAext}) for $n_f=5$ at three loops with only $\beta^2$ terms included, which still gave a good approximation, about half of one percent or better from the exact value, though of course not as good as the one discussed in this paper where $\beta^4$ terms are also included. 

Another possible extension is to include further exact results (in addition to the exact terms already present) for some color structures and/or other combinations of terms (when those are known) in the approximate expression. For example, at three loops we can include the full two-loop results in our expression and only have a small-$\beta$ expansion in $C^{(3)}$, i.e. we could consider the alternative expression $2 K_2 (\Gamma^{(2)}-K_2 \Gamma^{(1)})+C^{(3)}_{\beta^{2,4}}+K_3 \Gamma^{(1)}$. This, again, makes a negligible difference over the entire $\beta$ range, at the level of parts per million for much of it, with details depending on the number of flavors.

Our method is also clearly applicable to higher numbers of loops, and it could be utilized when the necessary information becomes available. For example, for a five-loop prediction, we would need to know the small-$\beta$ expansion of the cusp anomalous dimension at five loops as well as the result for the light-like $K_5$. 

\subsection{Further study of color structures}

We can also study the approximation separately for each color structure in the cusp anomalous dimension at each perturbative order.

At two loops, the $C_F C_A$ terms are not exact in $\Gamma_A^{(2)}$, as mentioned earlier, while the $C_F n_f$ terms are exact. Studying the approximation from asymptotics just for the $C_F C_A$ terms alone, we find excellent agreement with the exact result for those terms, better than one per mille everywhere in the $\beta$ range, and much smaller than that for most of the range. This is consistent with and expected from the excellence of the approximation for the total $\Gamma_A^{(2)}$.

At three loops, the $C_F^2 n_f$ and the $C_F n_f^2$ terms are exact in $\Gamma^{(3)}_A$, as mentioned earlier, but the $C_F C_A^2$ and $C_F C_A n_f$ terms are not exact. We study the approximation from asymptotics separately for those terms. We find excellent agreement with the exact result for both the $C_F C_A^2$ and $C_F C_A n_f$ terms, within a fraction of one per mille everywhere in the $\beta$ range, smaller than 0.1 per mille for the majority of the $\beta$ range, and smaller than one part per million at small speeds. This behavior is fully consistent with the behavior and excellence of the approximation for the total $\Gamma_A^{(3)}$.

At four loops, as mentioned earlier, the $C_F^3 n_f$, $C_F^2 n_f^2$, and $C_F n_f^3$ terms in $\Gamma^{(4)}_A$ are exact, but all the rest of the terms in $\Gamma^{(4)}_A$, i.e. the $C_F C_A^3$, $C_F^2 C_A n_f$, $C_F C_A^2 n_f$, $C_F C_A n_f^2$, $d_F d_F$, and $d_F d_A$ terms, are not exact. There exist exact results for some of these color structures, so one can make comparisons to them. The exact results for the $d_F d_F$ terms are very complicated \cite{BDHY}, but it is easier to make comparisons with the conjectured results for the $C_F^2 C_A n_f$ and $C_F C_A n_f^2$ terms \cite{GHKM1,GHKM2}. 

The $C_F^2 C_A n_f$ terms in $\Gamma^{(4)}$ are conjectured to be $2 K_3^{C_F n_f} (\Gamma^{(2)}-K_2 \Gamma^{(1)})+K_4^{C_F C_A n_f} \Gamma^{(1)}$ while the $C_F C_A n_f^2$ terms are conjectured to be $(19/81) n_f^2 T_F^2 (\Gamma^{(2)}-K_2 \Gamma^{(1)})+K_4^{C_A n_f^2} \Gamma^{(1)}$ \cite{GHKM1,GHKM2,AGG2}, where the superscripts in $K_3$ and $K_4$ denote the corresponding terms in them, and both of these conjectured expressions are consistent with the small-$\beta$ expansions in Eqs. (\ref{4lb2}) and (\ref{4lb4}) so they seem to be correct. We find superb agreement for both of these color structures between the conjectured results and our results from asymptotics. The difference is at the level of parts per million up to $\beta \approx 0.3$, less than 0.03 per mille for the vast majority of the $\beta$ range, and less than a small fraction of one per mille (0.3 per mille for $C_F^2 C_A n_f$, and 0.2 per mille for $C_F C_A n_f^2$) for all $\beta$. We note that $\beta^6$ terms are also available in the small-$\beta$ expansion for the $C_F C_A n_f^2$ terms \cite{BGHS}, but as can easily be seen from the above comparison there is negligible room for improvement.  

Furthermore, even though the $d_F d_F$ exact results \cite{BDHY} are very complicated, one can investigate further known terms of this color structure at small speeds \cite{BGHS}. Using the results in Ref. \cite{BGHS}, we find that the $\beta^6$ terms in the small-$\beta$ expansion of the $d_F d_F$ color structure at four loops are $\beta^6(-904/1225-10132 \, \zeta_2 /3675+53248 \, \zeta_3 /11025-718 \, \zeta_4 /735-2816 \, \zeta_5 /441+38944 \, \zeta_2 \zeta_3 /11025)$. Their contribution does not materially change the four-loop prediction: a difference of less than one part per million for much of the $\beta$ range, and everywhere less than 0.02 per mille for $n_f=3$, 0.05 per mille for $n_f=4$, and 0.7 per mille for $n_f=5$. Once again, this highlights the robustness of our approach and the reliability of our method.  

Finally, we can also investigate the effect of including the exact form of the conjectured  $C_F^2 C_A n_f$ and $C_F C_A n_f^2$ terms in our four-loop expression. Again, we find remarkable robustness in our method, consistent with all the previous checks. The difference between the results is negligible, of the order of parts per million for much of the $\beta$ range (with exact numbers depending on the number of flavors) and at the level of per mille for the entirety of the $\beta$ range. Thus, there can be no reasonable doubt that our four-loop result is very precise, and the inclusion of any future exact results or more terms in the small-$\beta$ expansion would make very little numerical difference.

\mysection{Conclusions}

An expression for the massive cusp anomalous dimension has been derived from its asymptotic behavior at small and large quark velocities through four loops. At two and three loops the expression predicts numerically the known exact results astonishingly well, and new calculations have been presented at four loops. The consistency and excellence of the results across different orders and number of flavors as well as color structures illustrates the success and robustness of the method. The expression is in general applicable to an arbitrary number of loops, so it can be utilized at five loops or higher once the small-$\beta$ behavior and the light-like cusp anomalous dimension are determined at those loops.

The method presented has been developed in terms of the quark velocity for the case of equal mass for the two eikonal lines, but the results can afterwards be reexpressed in terms of the cusp angle $\theta$; then, those results are valid for a given $\theta$ even when it describes cases with different masses for the two eikonal lines. Thus, the method is completely general and applies to any situation. The method can be readily extended to higher-term $\beta$ expansions as well as to higher loops once the necessary ingredients are known.

Calculations of soft anomalous dimensions, which are used in resummations for various processes, involve the cusp anomalous dimension as an essential component. Soft-gluon resummation has been very successful in approximating and predicting higher-order corrections for top-quark production and other heavy-quark processes and beyond. Thus, the derivation of highly accurate results for the cusp anomalous dimension at four loops is an important step towards more precise theoretical predictions for hard-scattering processes as well as a better understanding of the infrared behavior of QCD. 

\section*{Acknowledgements}
This material is based upon work supported by the National Science Foundation under Grant No. PHY 2112025.

\appendix

\mysection{Light-like cusp anomalous dimension}

The massless limit of the cusp anomalous dimension, 
i.e. the limit $\theta \rightarrow \infty$, can be written as
\beq
\lim_{\theta \rightarrow \infty} \Gamma^{(n)}= A^{(n)} \, \lim_{\theta \rightarrow \infty} \theta + R_n 
\label{lightlike}
\eeq
where $A^{(n)}=C_F K_n$ is the light-like cusp anomalous dimension,
and $C_F=(N_c^2-1)/2N_c$ with $N_c$ the number of colors.

At one loop $K_1=1$ and at two-loops \cite{KT82}
\beq
K_2=C_A \left(\frac{67}{36}-\frac{\zeta_2}{2}\right)-\frac{5}{9} n_f T_F \, ,
\label{K2}
\eeq
where $C_A=N_c$, $\zeta_2=\pi^2/6$, $T_F=1/2$, and $n_f$ is the number of light-quark flavors.

At three loops \cite{MVV04}
\beqa
K_3&=&C_A^2\left(\frac{245}{96}-\frac{67}{36}\zeta_2
+\frac{11}{24}\zeta_3+\frac{11}{8}\zeta_4 \right)
+C_F n_f T_F \left(-\frac{55}{48}+\zeta_3 \right)
\nonumber \\ &&
{}+C_A n_f T_F \left(-\frac{209}{216}+\frac{5}{9}\zeta_2
-\frac{7}{6}\zeta_3\right)-\frac{1}{27}n_f^2 T_F^2 \, ,
\label{K3}
\eeqa
with $\zeta_3=1.202056903 \cdots$ and $\zeta_4=\pi^4/90$.

At four loops  \cite{HKM,MPS}
\beqa
K_4 &=& C_A^3 \left(\frac{42139}{10368}-\frac{5525}{1296}\zeta_2
+\frac{1309}{432}\zeta_3+\frac{451}{64}\zeta_4
-\frac{451}{288}\zeta_5-\frac{313}{96}\zeta_6 
-\frac{11}{24}\zeta_2\zeta_3-\frac{\zeta_3^2}{16} \right) 
\nonumber \\ && \hspace{-3mm} 
{}+C_F^2 n_f T_F \left(\frac{143}{288}+\frac{37}{24} \zeta_3
-\frac{5}{2} \zeta_5\right)
+C_F C_A n_f T_F \left(-\frac{17033}{5184}+\frac{55}{48}\zeta_2
+\frac{29}{9}\zeta_3-\frac{11}{8}\zeta_4
+\frac{5}{4}\zeta_5-\zeta_2 \zeta_3 \right)
\nonumber \\ && \hspace{-3mm} 
{}+C_A^2 n_f T_F \left(-\frac{24137}{10368}
+\frac{635}{324} \zeta_2-\frac{361}{54}\zeta_3
-\frac{11}{24}\zeta_4+\frac{131}{72}\zeta_5+\frac{7}{6}\zeta_2\zeta_3 \right)
+C_F n_f^2 T_F^2 \left(\frac{299}{648}-\frac{10}{9} \zeta_3
+\frac{\zeta_4}{2} \right)
\nonumber \\ && \hspace{-3mm}  
{}+C_A n_f^2 T_F^2 \left(\frac{923}{5184}-\frac{19}{162}\zeta_2
+\frac{35}{27}\zeta_3-\frac{7}{12}\zeta_4\right)
+n_f^3 T_F^3 \left(-\frac{1}{81}+\frac{2}{27}\zeta_3 \right)
\nonumber \\ && \hspace{-3mm} 
{}+\frac{d_F^{abcd} d_F^{abcd}}{C_F N_c} n_f \left(\zeta_2 
-\frac{\zeta_3}{3}-\frac{5}{3}\zeta_5\right)
+\frac{d_F^{abcd} d_A^{abcd}}{C_F N_c} \left(-\frac{\zeta_2}{2}+\frac{\zeta_3}{6} 
+\frac{55}{12}\zeta_5-\frac{31}{8}\zeta_6-\frac{3}{2}\zeta_3^2\right) 
\label{K4}
\eeqa
where $\zeta_5=1.036927755\cdots$, $\zeta_6=\pi^6/945$, $d_F^{abcd} d_F^{abcd}/(C_F N_c)=(N_c^4-6N_c^2+18)/(48 N_c^2)$, and $d_F^{abcd} d_A^{abcd}/(C_F N_c)=N_c(N_c^2+6)/24$.

\mysection{Three-loop massive cusp anomalous dimension}

The cusp anomalous dimension at three loops is given by Eq. (\ref{cusp3loop}) with $C^{(3)}=C_F C_A^2 C^{' (3)}$, where $C^{' (3)}$ written as a function of $\beta$ is given by
\beqa
C^{' (3)}&=&-\frac{1}{2}+\frac{\zeta_2}{2}-\frac{\zeta_3}{2}
-\frac{9}{8}\zeta_4
+\frac{\zeta_2}{2}\ln\left(\frac{1-\beta}{1+\beta}\right)
-\frac{1}{4}\ln^2\left(\frac{1-\beta}{1+\beta}\right)
+\frac{1}{12}\ln^3\left(\frac{1-\beta}{1+\beta}\right)
-\frac{1}{24}\ln^4\left(\frac{1-\beta}{1+\beta}\right) 
\nonumber \\ && 
{}+\frac{1}{4}\ln^2\left(\frac{1-\beta}{1+\beta}\right)\ln \left(\frac{4\beta}{(1+\beta)^2}\right)
+\frac{3}{4} \ln\left(\frac{1-\beta}{1+\beta}\right) {\rm Li}_2\left(\frac{(1-\beta)^2}{(1+\beta)^2}\right)
-\frac{5}{8} {\rm Li}_3\left(\frac{(1-\beta)^2}{(1+\beta)^2}\right)
\nonumber \\ && \hspace{-18mm}
{}+\frac{(1+\beta^2)}{2\beta} \left\{-\frac{\zeta_3}{4}+\frac{15}{8}\zeta_4
-\left(\frac{\zeta_2}{2}-\frac{\zeta_3}{2}
+\frac{9}{8}\zeta_4\right) \ln\left(\frac{1-\beta}{1+\beta}\right)  
+\left(\frac{1}{4}+\zeta_2\right) \ln^2\left(\frac{1-\beta}{1+\beta}\right)
\right.
\nonumber \\ && \quad  
{}-\left(\frac{1}{12}+\frac{\zeta_2}{3}\right)\ln^3\left(\frac{1-\beta}{1+\beta}\right)
+\frac{7}{24}\ln^4\left(\frac{1-\beta}{1+\beta}\right)
-\frac{1}{24}\ln^5\left(\frac{1-\beta}{1+\beta}\right)
\nonumber \\ && \quad   
{}+\frac{1}{2}\ln^2\left(\frac{1-\beta}{1+\beta}\right) \ln \left(\frac{4\beta}{(1+\beta)^2}\right)
-\frac{1}{2}\ln^3\left(\frac{1-\beta}{1+\beta}\right) \ln \left(\frac{4\beta}{(1+\beta)^2}\right)
\nonumber \\ && \quad   
{}-\frac{3}{4} \ln^2\left(\frac{1-\beta}{1+\beta}\right) {\rm Li}_2\left(\frac{(1-\beta)^2}{(1+\beta)^2}\right)
+\frac{1}{4} {\rm Li}_2\left(\frac{4\beta}{(1+\beta)^2}\right)
+\frac{1}{4} {\rm Li}_3\left(\frac{(1-\beta)^2}{(1+\beta)^2}\right)
\nonumber \\ && \quad    \left.
{}+\frac{7}{4} \ln\left(\frac{1-\beta}{1+\beta}\right) {\rm Li}_3\left(\frac{(1-\beta)^2}{(1+\beta)^2}\right)
+\frac{1}{2} {\rm Li}_3\left(\frac{4\beta}{(1+\beta)^2}\right)-\frac{15}{8} {\rm Li}_4\left(\frac{(1-\beta)^2}{(1+\beta)^2}\right) \right\}
\nonumber \\ && \hspace{-18mm}
+\frac{(1+\beta^2)^2}{4\beta^2} \left\{-\frac{19}{8}\zeta_4+\frac{3}{2}\zeta_5-\frac{\zeta_2 \zeta_3}{2}
-\left(\frac{3}{2} \zeta_3-\frac{15}{8} \zeta_4 \right) \ln\left(\frac{1-\beta}{1+\beta}\right)
-\left(\zeta_2-\frac{\zeta_3}{4}\right) \ln^2\left(\frac{1-\beta}{1+\beta}\right)
\right.
\nonumber \\ && \quad  
{}+\frac{2}{3} \zeta_2 \ln^3\left(\frac{1-\beta}{1+\beta}\right)
-\frac{1}{4}\ln^4\left(\frac{1-\beta}{1+\beta}\right)+\frac{11}{120}\ln^5\left(\frac{1-\beta}{1+\beta}\right)
\nonumber \\ && \quad  
{}+\ln\left(\frac{4\beta}{(1+\beta)^2}\right)\left[\zeta_3+\zeta_2 \ln\left(\frac{1-\beta}{1+\beta}\right)
-\zeta_2 \ln^2\left(\frac{1-\beta}{1+\beta}\right) \right.
\nonumber \\ && \hspace{37mm} \left.
{}+\frac{1}{3}\ln^3\left(\frac{1-\beta}{1+\beta}\right)-\frac{1}{6}\ln^4\left(\frac{1-\beta}{1+\beta}\right)\right]
\nonumber \\ && \quad 
{}-\ln^2\left(\frac{1-\beta}{1+\beta}\right) \ln^2\left(\frac{4\beta}{(1+\beta)^2}\right)
+\ln\left(\frac{1-\beta}{1+\beta}\right)  \ln^3\left(\frac{4\beta}{(1+\beta)^2}\right)
-\frac{1}{8} \ln^4\left(\frac{4\beta}{(1+\beta)^2}\right)
\nonumber \\ && \quad
{}+\left[\frac{\zeta_2}{2}-\zeta_2\ln\left(\frac{1-\beta}{1+\beta}\right)
-2\ln^2\left(\frac{1-\beta}{1+\beta}\right)
-\frac{1}{12}\ln^3\left(\frac{1-\beta}{1+\beta}\right) \right.
\nonumber \\ && \hspace{13mm} \left.
{}+\ln\left(\frac{1-\beta}{1+\beta}\right) 
\ln\left(\frac{4\beta}{(1+\beta)^2}\right) \right]
{\rm Li}_2\left(\frac{(1-\beta)^2}{(1+\beta)^2}\right)
\nonumber \\ && \quad 
{}-\frac{1}{4} {\rm Li}_2^2\left(\frac{(1-\beta)^2}{(1+\beta)^2}\right)
+\frac{1}{2} \ln^2\left(\frac{4\beta}{(1+\beta)^2}\right) {\rm Li}_2\left(\frac{4\beta}{(1+\beta)^2}\right)
\nonumber \\ && \quad 
{}+\frac{1}{4} {\rm Li}_2^2\left(\frac{4\beta}{(1+\beta)^2}\right)
-\frac{1}{2} \ln^2\left(\frac{4\beta}{(1-\beta)^2}\right)
{\rm Li}_2\left(\frac{-(1-\beta)^2}{4\beta}\right) 
\nonumber \\ && \quad 
{}+\left[\frac{\zeta_2}{2}+\frac{3}{2}\ln\left(\frac{1-\beta}{1+\beta}\right)  
-\frac{1}{4}\ln^2\left(\frac{1-\beta}{1+\beta}\right)
-\ln\left(\frac{4\beta}{(1+\beta)^2}\right)\right]{\rm Li}_3\left(\frac{(1-\beta)^2}{(1+\beta)^2}\right)
\nonumber \\ && \quad 
{}+\left[\ln\left(\frac{1-\beta}{1+\beta}\right)
-\ln\left(\frac{4\beta}{(1+\beta)^2}\right)\right] 
{\rm Li}_3\left(\frac{4\beta}{(1+\beta)^2}\right)
\nonumber \\ && \quad 
{}+\left[2\ln\left(\frac{1-\beta}{1+\beta}\right)
-\ln \left(\frac{4\beta}{(1+\beta)^2}\right) \right] 
{\rm Li}_3\left(\frac{-(1-\beta)^2}{4\beta}\right)+\frac{9}{8} \ln\left(\frac{1-\beta}{1+\beta}\right)  {\rm Li}_4\left(\frac{(1-\beta)^2}{(1+\beta)^2}\right)
\nonumber \\ && \quad  \left.
{}+{\rm Li}_4\left(\frac{4\beta}{(1+\beta)^2}\right)-{\rm Li}_4\left(\frac{-(1-\beta)^2}{4\beta}\right)-\frac{3}{2} {\rm Li}_5\left(\frac{(1-\beta)^2}{(1+\beta)^2}\right) \right\}
\nonumber \\ && \hspace{-18mm} 
{}+\frac{(1+\beta^2)^3}{32\beta^3} \left\{-3 \zeta_5
-4 \zeta_4 \ln\left(\frac{1-\beta}{1+\beta}\right)
-3 \zeta_3 \ln^2\left(\frac{1-\beta}{1+\beta}\right)
-\frac{4}{3} \zeta_2 \ln^3\left(\frac{1-\beta}{1+\beta}\right)
-\frac{1}{5} \ln^5\left(\frac{1-\beta}{1+\beta}\right)\right.
\nonumber \\ && \quad \; 
{}-\frac{2}{3} \ln^3\left(\frac{1-\beta}{1+\beta}\right) {\rm Li}_2\left(\frac{(1-\beta)^2}{(1+\beta)^2}\right) 
+\ln^2\left(\frac{1-\beta}{1+\beta}\right){\rm Li}_3\left(\frac{(1-\beta)^2}{(1+\beta)^2}\right)   
\nonumber \\ && \quad \; 
{}-2 \ln\left(\frac{1-\beta}{1+\beta}\right) {\rm Li}_4\left(\frac{(1-\beta)^2}{(1+\beta)^2}\right)
+3 {\rm Li}_5\left(\frac{(1-\beta)^2}{(1+\beta)^2}\right) 
\nonumber \\ && \quad \;  \left.
{}+H_{1,1,0,0,1}\left(\frac{4\beta}{(1+\beta)^2}\right)+H_{1,0,1,0,1}\left(\frac{4\beta}{(1+\beta)^2}\right) \right\}
\nonumber \\ && \hspace{-18mm}
{}+\frac{(1-\beta^2)}{16\beta}
\left\{-2 \zeta_2 \zeta_3 - 2 \zeta_3 \ln\left(\frac{1-\beta}{1+\beta}\right) \ln\left(\frac{1+\beta}{2}\right)
+\left[\frac{3}{2} \zeta_4 -\frac{1}{6} \ln^4\left(\frac{1-\beta}{1+\beta}\right)\right] \ln\beta \right.
\nonumber \\ && \quad  
{}+2 \zeta_3 \left[{\rm Li}_2\left(\frac{-1+\beta}{1+\beta}\right)
+{\rm Li}_2\left(\frac{2\beta}{1+\beta}\right)\right]
-\frac{2}{3} \ln^3\left(\frac{1-\beta}{1+\beta}\right)
\left[{\rm Li}_2\left(\frac{1-\beta}{1+\beta}\right)
-{\rm Li}_2\left(\frac{-1+\beta}{1+\beta}\right)\right]
\nonumber \\ && \quad 
{}+2 \ln^2\left(\frac{1-\beta}{1+\beta}\right) \left[{\rm Li}_3\left(\frac{1-\beta}{1+\beta}\right)
-{\rm Li}_3\left(\frac{-1+\beta}{1+\beta}\right)\right]
\nonumber \\ && \quad 
{}-4 \ln\left(\frac{1-\beta}{1+\beta}\right) 
\left[{\rm Li}_4\left(\frac{1-\beta}{1+\beta}\right)
-{\rm Li}_4\left(\frac{-1+\beta}{1+\beta}\right)\right]
+4 {\rm Li}_5\left(\frac{1-\beta}{1+\beta}\right)
-4 {\rm Li}_5\left(\frac{-1+\beta}{1+\beta}\right)
\nonumber \\ && \quad 
{}+4 H_{1,0,1,0,0}\left(\frac{1-\beta}{1+\beta}\right)+4 H_{-1,0,1,0,0}\left(\frac{1-\beta}{1+\beta}\right)
\nonumber \\ && \quad \left.
{}-4 H_{1,0,-1,0,0}\left(\frac{1-\beta}{1+\beta}\right)-4 H_{-1,0,-1,0,0}\left(\frac{1-\beta}{1+\beta}\right) \right\} \, ,
\label{C3p}
\eeqa
where explicit expressions for the six distinct weight-five harmonic polylogarithms $H$ in the above equation can be found in the Appendix of Ref. \cite{NK3loopcusp}.


\begin{thebibliography}{99}

\bibitem{AMP}
A.M. Polyakov, {\sl Gauge fields as rings of glue}, Nucl. Phys. B {\bf 164}, 171 (1980).

\bibitem{BNS}
R.A. Brandt, F. Neri, and M. Sato, {\sl Renormalization of loop functions for all loops}, Phys. Rev. D {\bf 24}, 879 (1981).

\bibitem{KnSc}
D. Knauss and K. Scharnhorst, {\sl Two-loop renormalization of non-smooth string operators in Yang-Mills theory}, Annalen Phys. {\bf 41}, 331 (1984).

\bibitem{IKR}
S.V. Ivanov, G.P. Korchemsky, and A.V. Radyushkin, {\sl The infrared asymptotic behavior of perturbative QCD. Contour gauges}, Yad. Fiz. {\bf 44}, 230 (1986) [Sov. J. Nucl. Phys. {\bf 44}, 145 (1986)].  

\bibitem{KRsj}
G.P. Korchemsky and A.V. Radyushkin, {\sl Infrared asymptotics of perturbative QCD: renormalization properties of the Wilson loops in higher orders of perturbation theory}, Yad. Fiz. {\bf 44}, 1351 (1986) [Sov. J. Nucl. Phys. {\bf 44}, 877 (1986)].

\bibitem{KR1}
G.P. Korchemsky and A.V. Radyushkin, {\sl Loop-space formalism and renormalization group for the infrared asymptotics of QCD}, Phys. Lett. B {\bf 171}, 459 (1986). 

\bibitem{KR2}
G.P. Korchemsky and A.V. Radyushkin, {\sl Renormalization of the Wilson loops beyond the leading order}, Nucl. Phys. B {\bf 283}, 342 (1987).

\bibitem{GPK}
G.P. Korchemsky, {\sl Asymptotics of the Altarelli-Parisi-Lipatov evolution kernels of parton distributions}, Mod. Phys. Lett. A {\bf 4}, 1257 (1989).

\bibitem{KR3}
G.P. Korchemsky and A.V. Radyushkin, {\sl Infrared factorization, Wilson lines and the heavy quark limit}, Phys. Lett. B {\bf 279}, 359 (1992) [hep-ph/9203222].

\bibitem{KMM}
W. Kilian, P. Manakos, and T. Mannel, {\sl Leading and subleading logarithmic QCD corrections to bilinear heavy quark currents}, Phys. Rev. D {\bf 48}, 1321 (1993).

\bibitem{AGhq}
A.G. Grozin, {\sl Heavy quark effective theory}, Springer Tracts Mod. Phys., Vol. {\bf 201} (Springer, 2004). 

\bibitem{NK2loop}
N. Kidonakis, {\sl Two-loop soft anomalous dimensions and next-to-next-to-leading-logarithm resummation for heavy quark production}, Phys. Rev. Lett. {\bf 102}, 232003 (2009) [arXiv:0903.2561].

\bibitem{NK2lp}
N. Kidonakis, {\sl Two-loop soft anomalous dimensions with massive and massless quarks}, in Proceedings of the DPF-2009 Conference, eConf C090726 [arXiv:0910.0473].

\bibitem{NKtW}
N. Kidonakis, {\sl Two-loop soft anomalous dimensions for single top quark associated production with a $W^-$ or $H^-$}, Phys. Rev. D {\bf 82}, 054018 (2010) [arXiv:1005.4451]. 

\bibitem{CHMS}
D. Correa, J. Henn, J. Maldacena, and A. Sever, {\sl The cusp anomalous dimension at three loops and beyond}, JHEP {\bf 05}, 098 (2012) [arXiv:1203.1019].

\bibitem{GHKM1}
A. Grozin, J.M. Henn, G.P. Korchemsky, and P. Marquard, {\sl Three-loop cusp anomalous dimension in QCD}, Phys. Rev. Lett. {\bf 114}, 062006 (2015) [arXiv:1409.0023]. 

\bibitem{GHKM2}
A. Grozin, J.M. Henn, G.P. Korchemsky, and P. Marquard, {\sl The three-loop cusp anomalous dimension in QCD and its supersymmetric extensions}, JHEP {\bf 01}, 140 (2016) [arXiv:1510.07803].

\bibitem{NK3loopcusp}
N. Kidonakis, {\sl Three-loop cusp anomalous dimension and a conjecture for $n$ loops}, Int. J. Mod. Phys. A {\bf 31}, 1650076 (2016) [arXiv:1601.01666].

\bibitem{GHS}
A.G. Grozin, J.M. Henn, and M. Stahlhofen, {\sl On the Casimir scaling violation in the cusp anomalous dimension at small angle}, JHEP {\bf 10}, 052 (2017) [arXiv:1708.01221].

\bibitem{AGG1}
A. Grozin, {\sl Four-loop cusp anomalous dimension in QED}, JHEP {\bf 06}, 073 (2018) [arXiv:1805.05050].

\bibitem{NK3loop}
N. Kidonakis, {\sl Soft anomalous dimensions for single-top production at three loops}, Phys. Rev. D {\bf 99}, 074024 (2019) [arXiv:1901.09928].

\bibitem{BGHS}
R. Bruser, A.G. Grozin, J.M. Henn, and M. Stahlhofen, {\sl Matter dependence of the four-loop QCD cusp anomalous dimension: from small angles to all angles}, JHEP {\bf 05}, 186 (2019) [arXiv:1902.05076].

\bibitem{BDHY}
R. Bruser, C. Dlapa, J.M. Henn, and K. Yan, {\sl Full angle dependence of the four-loop cusp anomalous dimension in QED}, Phys. Rev. Lett. {\bf 126}, 021601 (2021) [arXiv:2007.04851].

\bibitem{GLP}
A.G. Grozin, R.N. Lee, and A.F. Pikelner, {\sl Four-loop QCD cusp anomalous dimension at small angle}, JHEP {\bf 11}, 094 (2022) [arXiv:2208.09277].

\bibitem{AGG2}
A. Grozin, {\sl QCD cusp anomalous dimension: current status}, arXiv:2212.05290.

\bibitem{NKGS1}
N. Kidonakis and G. Sterman, {\sl Subleading logarithms in QCD hard scattering}, Phys. Lett. B {\bf 387}, 867 (1996). 

\bibitem{NKGS2}
N. Kidonakis and G. Sterman, {\sl Resummation for QCD hard scattering}, Nucl. Phys. B {\bf 505}, 321 (1997) [hep-ph/9705234].

\bibitem{KOS}
N. Kidonakis, G. Oderda, and G. Sterman, {\sl Evolution of color exchange in QCD hard scattering}, Nucl. Phys. B {\bf 531}, 365 (1998) [hep-ph/9803241].

\bibitem{NKtt2l}
N. Kidonakis, {\sl Next-to-next-to-leading soft-gluon corrections for the top quark cross section and transverse momentum distribution}, Phys. Rev. D {\bf 82}, 114030 (2010) [arXiv:1009.4935].

\bibitem{FK2020}
M. Forslund and N. Kidonakis, {\sl Resummation for $2 \to n$ processes in single-particle-inclusive kinematics}, Phys. Rev. D {\bf 102}, 034006 (2020) [arXiv:2003.09021].

\bibitem{NKuni}
N. Kidonakis, {\sl Soft anomalous dimensions and resummation in QCD}, Universe {\bf 6}, 165 (2020) [arXiv:2008.09914].

\bibitem{KT82}
J. Kodaira and L. Trentadue, {\sl Summing soft emission in QCD}, Phys. Lett. {\bf 112B}, 66 (1982)

\bibitem{MVV04}
S. Moch, J.A.M. Vermaseren, and A. Vogt, {\sl The three-loop splitting functions in QCD: the non-singlet case}, Nucl. Phys. B {\bf 688}, 101 (2004) [arXiv:hep-ph/0403192].

\bibitem{HKM}
J.M. Henn, G.P. Korchemsky, and B. Mistlberger, {\sl The full four-loop cusp anomalous dimension in $N=4$ super Yang-Mills and QCD}, JHEP {\bf 04}, 018 (2020) [arXiv:1911.10174].

\bibitem{MPS}
A. von Manteuffel, E. Panzer, and R.M. Schabinger, {\sl Cusp and collinear anomalous dimensions in four-loop QCD from form factors}, Phys. Rev. Lett. {\bf 124}, 162001 (2020) [arXiv:2002.04617].

\end{thebibliography}
\end{document}